\documentclass[aps,prd,superscriptaddress,nofootinbib,amsmath,amsfonts,preprintnumbers,groupedaddress,showpacs,10pt,english]{revtex4-1}
\usepackage{amsmath}
\usepackage{amssymb}
\usepackage{babel}
\usepackage{wrapfig}
\usepackage{cancel}
\usepackage{bigints}
\newcommand{\nn}{\nonumber \\}
\newcommand{\e}{\mathrm{e}}
\usepackage{relsize,exscale}
\makeatletter

\usepackage{array,multirow,graphicx}
\usepackage{dcolumn}
\usepackage{newlfont}
\usepackage{bm}
\usepackage[colorlinks,citecolor=blue,urlcolor=blue,linkcolor=blue]{hyperref}
\usepackage[figtopcap]{subfigure}
\usepackage{color}

\allowdisplaybreaks[4]

\begin{document}

\tolerance=5000

\title{Black holes with Lagrange multiplier and mimetic potential \\
in mimetic gravitational theory: multi-horizon black holes}

\author{G.~G.~L.~Nashed}
\email{nashed@bue.edu.eg}
\affiliation {Centre for Theoretical Physics, The British University, P.O. Box
43, El Sherouk City, Cairo 11837, Egypt}
\author{Shin'ichi~Nojiri}
\email{nojiri@gravity.phys.nagoya-u.ac.jp}
\affiliation{Department of Physics, Nagoya University, Nagoya 464-8602,
Japan \\
\& \\
Kobayashi-Maskawa Institute for the Origin of Particles and the Universe,
Nagoya University, Nagoya 464-8602, Japan }


\date{}

\begin{abstract}
In this paper, we employ the mimetic field equations coupled with the Lagrange multiplier and mimetic potential to derive non-trivial spherically symmetric black hole (BH) solutions.
We divided this study into three cases: The first one in which we take the Lagrange multiplier and mimetic potential to have vanishing value and derive a BH solution
that completely coincides with the BH of the Einstein general relativity despite the non-vanishing value of the mimetic field.
The first case is completely consistent with the previous studies in the literature that mimetic theory coincides with GR \cite{Nashed:2018qag}.
In the second case, we derive a solution with a constant value of the mimetic potential and a dynamical value of the Lagrange multiplier.
This solution has no horizon and therefore the obtained spacetime does not correspond to the BH.
In this solution, there appears the region of the Euclidian signature where the signature of the diagonal
components of the metric is $(+,+,+,+)$ or the region with two times where the signature is $(+,+,-,-)$.
Finally, we derive a BH solution with non-vanishing values of the Lagrange multiplier, mimetic potential, and mimetic field.
This BH shows a soft singularity compared with the Einstein BH solution.
The relevant physics of the third case is discussed by showing their behavior of the metric potential at infinity,
calculating their energy conditions, and study their thermodynamical quantities.
We give a brief discussion on how our third case can generate a BH with three horizons
as in the de Sitter-Reissner-Nordstr\"om black hole spacetime, where the largest horizon is the cosmological one
and two correspond to the outer and inner horizons of the BH.
Even in the third case, there appears the region of the Euclidian signature or the region with two times.
We give a condition that such unphysical region(s) is hidden inside the black hole horizon and the existence of the region(s) becomes less unphysical.
We also study the thermodynamics of the multi-horizon BH and consider the extremal case, where the radii of two horizons coincide with each other.
We observe that the haking temperature and the heat capacity vanish in the extremal limit.
Finally, we would like to stress the fact that in spite that the field equations we use have no cosmological constant our
BH solutions of the second and third case behave as AdS/dS.
\end{abstract}


\maketitle
\section{Introduction}\label{S1}

The key of diffeomorphism invariance in the Einstein general relativity is to stimulate the redundancy in the presentation
of the dynamical degrees of freedom in exchange for improving the simplicity and the elegancy in the formulation.
The metric $g_{\alpha \beta}$ that has ten components are employed to describe two dynamical degrees of freedom for the graviton field.
Thus, it is natural to search for an amended gravitational theory without enlarging the degrees of freedom of the gravitational system when keeping
the diffeomorphism invariance.
A few years ago, this amended gravitational theory was constructed in \cite{Chamseddine:2013kea}, using the idea of rolling
the dynamical metric $g_{\alpha \beta}$ which depends on an auxiliary metric $\bar{g}_{\alpha\beta}$.

Mimetic gravitational theory is considered as one of the most attractive theories of gravity, which without inserting any extra matter field,
representing the dark piece of the universe that is representing as a geometrical effect \cite{Chamseddine:2013kea}.
In the mimetic theory, the conformal degree of freedom of gravitational field is isolated by inserting the relation between
the physical metric $g_{\alpha \beta}$, the auxiliary metric $\bar{g}_{\alpha \beta}$ and a mimetic field which is the scalar field as:
\begin{equation}
\label{trans1}
g_{\alpha\beta}=\mp \left(\bar{g}^{\mu \nu} \partial_\mu \zeta\partial_\nu \zeta \right) \bar{g}_{\alpha\beta}\,.
\end{equation}
Here $\bar{g}^{\mu \nu}$ is the inverse of $\bar{g}_{\mu \nu}$.
Equation~(\ref{trans1}) implies that the mimetic field should yield:
\begin{equation}
\label{trans2}
g^{\alpha \beta}\partial_\alpha \zeta \partial_\beta \zeta= \mp 1\,.
\end{equation}
Therefore, $\partial_\alpha \zeta$ is timelike and spacelike (the signature of $g_{\alpha \beta}$ is chosen in this study as
$\left( g_{\mu\nu} \right) = \mathrm{diag} \left(+, -, -, - \right)$)
when we consider the positive and negative signs in (\ref{trans1}) or (\ref{trans2}), respectively.

The field equations of the gravitational action are equal to the one which can derive from the action written in terms of the physical metric
with the prescription of restriction (\ref{trans2}), through the use of the Lagrange multiplier \cite{Chamseddine:2013kea}.
The conformal degree of freedom is dynamical quantity even in the absence
of matter, and this imitates the situation of cold dark matter evolution of our universe in the background.
Moreover, it has been explained that the scalar field can imitate the gravitational behavior of any configuration of matter \cite{Chamseddine:2014vna,Lim:2010yk}.
Thereafter, the mimetic model was expanded to the studies of inflation, theories with non-singular
cosmological, dark energy, and black hole solutions \cite{Chamseddine:2014vna,Chamseddine:2016uef,Chamseddine:2016ktu,
HosseiniMansoori:2020mxj,Mirzagholi:2014ifa,Myrzakulov:2015kda,Arroja:2015yvd,Sebastiani:2016ras,Dutta:2017fjw,
Saadi:2014jfa,Firouzjahi:2018xob,Gorji:2020ten,Matsumoto:2015wja,
Momeni:2015aea,Astashenok:2015qzw,Sadeghnezhad:2017hmr,Nozari:2019esz,Solomon:2019qgf,Shen:2019nyp,Ganz:2019vre,
deCesare:2019pqj,Nozari:2019shm,deCesare:2018cts,Ganz:2018mqi,Nashed:2021pkc,Nashed:2021ctg,
Ganz:2018vzg,Sheykhi:2020dkm,Sheykhi:2020fqf,Nojiri:2014zqa,Astashenok:2015haa,Nojiri:2016ppu,Nashed:2018urj,
Nashed:2018qag,Nashed:2018aai,Nojiri:2017ygt,Nojiri:2016vhu,Odintsov:2018ggm,Casalino:2018wnc}.
Moreover, it was argued that in the four-dimensional Einstein-Maxwell theory and for asymptotically
AdS spacetime, there exist BH solutions whose event horizons
could have zero or negative constant curvature and therefore, their topologies are no longer
the two-sphere $S^2$ \cite{Lemos:1994xp,Cai:1996eg,Brill:1997mf,Cai:1997ii,Cai:1998vy,Hajkhalili:2018thm,Sheykhi:2007gw,Sheykhi:2006cw}.
Later, many modifications of mimetic gravity have been established, $f(R)$ mimetic gravity, mimetic gravity with the Lagrange multipliers \cite{Cid:2012dh,Capozziello:2010uv}.
Moreover, the presence of mimetic potential and the Lagrange multiplier supported the
possibility for recognizing different cosmologies, \cite{Odintsov:2015cwa,Odintsov:2015cwa}.
Besides that, it has been proved that the original setting of the mimetic theory forecasts that gravitational wave
(GW) propagates at the speed of light, confirming agreement with the results of the event GW170817 and its optical
counterpart \cite{Casalino:2018tcd,Casalino:2018wnc}.
The mimetic gravity is also generalized to the
$f(R)$ mimetic gravity \cite{Nojiri:2014zqa,Odintsov:2015wwp,Oikonomou:2016pkp,Oikonomou:2016fxb,Oikonomou:2015lgy,
Myrzakulov:2016hrx,Odintsov:2015cwa,Odintsov:2016imq,Odintsov:2016oyz,Nojiri:2017ygt,Odintsov:2018ggm,Chen:2020zzs}
and the Gauss-Bonnet mimetic gravity \cite{Astashenok:2015haa,Oikonomou:2015qha,Zhong:2016mfv,Zhong:2018tqn}.
Especially, a unified formalism of early inflation and
late-time acceleration in the frame of the mimetic $f(R)$ gravity was also constructed in \cite{Nojiri:2016vhu}
where the authors confirmed the inflationary era in contrast to the $f(R)$ gravity.

The exact solution is an ingredient role in GR and its modifications.
This is because that the thermodynamics and dynamics of the gravitational model are often referred to as the presence
of an exact solution, which solves the field equations. Moreover, to introduce a
new technique to find the exact solutions is also an important and natural way to construct modified gravity \cite{Stephani:2003tm}.
It is the aim of this study we are going to consider mimetic gravity theory, in which we will fix the Lagrange multiplier, mimetic potential, mimetic field,
and the unknown of the metric potentials.

The structure of this research is as follows: In Section~\ref{S2}, we give the cornerstone of the mimetic gravity coupled with mimetic potential and Lagrange multiplier.
Also in Section~\ref{S2}, we apply the field equations of mimetic theory coupled to the Lagrange multiplier and mimetic potential
to a spherically symmetric spacetime that has two unknown functions and derives the non-linear differential equations which have five unknown functions.
We solve these non-linear differential equations under three different constraints related to the mimetic potential and the  Lagrange multiplier and derive
BH solutions which are different from the BH in GR.
The invariants of these BHs are calculated in Section~\ref{S2} and we show that the singularity of our BHs is much milder compared with GR BHs.
In the BH solutions we find, there is a case that the spacetime has three horizons which may be the cosmological horizon and the inner and outer BH horizons.
Furthermore in the spacetime, there appears the region of the Euclidian signature where the signature of the diagonal
components of the metric is $(+,+,+,+)$ or the region with two times where the signature is $(+,+,-,-)$.
We find a condition that the region(s) is hidden inside the black hole horizon and the existence of the region(s) becomes less unphysical.
We also study the thermodynamics of the multi-horizon BH.
In Section~\ref{S3}, we study the energy conditions of these BH and show that when the mimetic potential has a constant value the strong energy condition
is not satisfied while when mimetic potential and the Lagrange multiplier have non-trivial forms all the energy conditions are satisfied.
In Section~\ref{S4}, we study the thermodynamics of the different BHs derived in this study and show that all the thermodynamics of the BHs
are satisfied and consistent with the results presented in the literature \cite{Nashed:2021ctg,Nashed:2021xtt,Nashed:2021rgn,Nashed:2021sey}.
In Section~\ref{S6}, we study the stability of the derived BHs using the procedure of geodesic deviations.
We study the conditions of stability analytically and graphemically.
We summarized the results of this study in the final Section~\ref{S66}.

\section{Spherically symmetric BH solutions in mimetic gravity coupled with mimetic potential and Lagrange multiplier}\label{S2}

The expression of ``mimetic dark matter" was presented in the literature by Mukhanov and Chamseddine \cite{Chamseddine:2014vna},
in spite that such theories had already been discussed in \cite{Lim:2010yk,Gao:2010gj,Capozziello:2010uv,Sebastiani:2016ras}.

In this study, we are going to present the mimetic gravity coupled with mimetic potential and the Lagrange multiplier.
The gravitational action of the mimetic gravity coupled with the Lagrange multiplier $\lambda$ and mimetic potential $V(\zeta)$
takes the following form \cite{Nojiri:2014zqa}:
\begin{equation}
\label{actionmimeticfraction}
S=\int \mathrm{d}x^4\sqrt{-g}\left\{ R \left(g_{\mu \nu}\right)
 -V(\zeta)+\lambda \left(g^{\mu \nu}\partial_{\mu}\zeta\partial_{\nu}\zeta
+1\right)\right \}+L_\mathrm{matt}\, ,
\end{equation}
with $L_\mathrm{matt}$ being the Lagrangian of the matter fluid present.
In this paper, we choose the gravitational coupling and the light speed to be unity.
Varying the action (\ref{actionmimeticfraction}) w.r.t. the metric tensor $g_{\mu \nu}$, we get the following field equations:
\begin{align}
\label{aeden}
& 0=R_{\mu \nu}-\frac{1}{2}g_{\mu \nu}R
+\frac{1}{2}g_{\mu \nu}\left\{\lambda \left( g^{\rho
\sigma}\partial_{\rho}\zeta\partial_{\sigma}\zeta+1\right) -V(\zeta)\right\}-\lambda
\partial_{\mu}\zeta \partial_{\nu}\zeta +\frac{1}{2}T_{\mu \nu} \, ,
\end{align}
where $T_{\mu \nu}$ is the energy momentum tensor corresponding to the matter fluids present.
In this study, we put $T_{\mu \nu}=0$.
Additionally, varying the action w.r.t. the auxiliary scalar field $\zeta$, we get
\begin{equation}\label{scalvar}
2\nabla^{\mu} (\lambda \partial_{\mu}\zeta)+V'(\zeta)=0\, ,
\end{equation}
where the ``prime'' refers to the differentiation w.r.t. the auxiliary scalar $\zeta$.
Finally, varying the action (\ref{actionmimeticfraction}) w.r.t. the Lagrange multiplier $\lambda$, we get
\begin{equation}\label{lambdavar}
g^{\rho \sigma}\partial_{\rho}\zeta \partial_{\sigma}\zeta=-1\, ,
\end{equation}
and as it can be seen, the above equation is identical to Eq.~(\ref{trans2}).

Now, we are going to apply the field equations of the mimetic theory, (\ref{aeden}), and (\ref{scalvar}) to a spherically symmetric spacetime
that has two unknown functions which has the following form:
\begin{equation}
\label{met}
ds^{2}=h(r)dt^{2}-\frac{dr^{2}}{h_1(r)}-r^2\left(d\theta^{2}+r^2\sin^2\theta d\phi^{2}\right)\,,
\end{equation}
where $h(r)$ and $h_1(r)$ are unknown functions, which we will determine by using the field equations.
We also suppose that the auxiliary scalar field depends only on the radial coordinate $r$.
Applying the field equations (\ref{aeden}) and (\ref{scalvar}) to spacetime (\ref{met}),
we obtain the following nonlinear differential equations:

\noindent
The $(t,t)$-component of the field equation (\ref{aeden}) is:
\begin{align}
\label{eqtt}
0=&2 - 2h_1 - 2rh'_1 +V r^2\,,
\end{align}
the $(r,r)$-component of the field equation (\ref{aeden}) is:
\begin{align}
\label{eqrr}
0=&2h - 2h_1 h'r - 2h h_1-2\lambda \zeta'^2 h_1 h r^2 +Vh r^2 \,,
\end{align}
both of the $(\theta,\theta)$ and $(\phi,\phi)$-components of the field equation (\ref{aeden}) have the form:
\begin{align}
\label{fe}
0=& h'^2h_1 r - 2h_1 h'h -2 h'_1 h^2 - h' h'_1h r - 2h''h_1 h r + 2V h^2r \,,
\end{align}
and the field equation (\ref{scalvar}) takes the form:
\begin{equation}
\label{chr}
0=\lambda \zeta'^2 h_1 h' r - V'h r+2 \lambda' \zeta'^2 h_1h r + 2\lambda \zeta''h_1 \zeta' h r + \lambda \zeta'^2h'_1 h r
+ 4 \lambda \zeta'^2 h_1 h \,,
\end{equation}
where $h\,\equiv h(r)$, $h_1\,\equiv h_1(r)$, $V\,\equiv V(r)$, $\lambda\,\equiv \lambda(r)$, $h'=\frac{dh}{dr}$, $h'_1=\frac{dh_1}{dr}$,
$V'=\frac{dV}{dr}$, $\zeta'=\frac{d\zeta}{dr}$, and $\lambda'=\frac{d\lambda}{dr}$.
We will solve the aforementioned nonlinear differential equations, i.e., (\ref{fe}) and (\ref{chr})
for the following three different cases:

\

\noindent
\underline{Case I: $V(r)=\lambda(r)=0$:}\vspace{0.2cm}\\
When $V(r)=\lambda(r)=0$, the analytic solution of the nonlinear differential equations, (\ref{fe}) and (\ref{chr}) takes the following forms
\begin{align}
\label{sol1}
h(r)=&h_1(r)= 1+\frac{2C}{r}=1-\frac{2M}{r}\,, \qquad \qquad C=-M\,,\nonumber\\
\zeta(r)=& \sqrt {r^2-2\,rM}+M\ln \left(r -M+\sqrt {r}\sqrt {r-2\,M}\right)\,.
\end{align}
Equation~(\ref{sol1}) shows that when the mimetic potential and the Lagrange multiplier are vanishing, we return to the well-known BH of GR, i.e.,
the Schwarzschild solution.
It is of interest to note that the mimetic field of Eq.~(\ref{sol1}) satisfy the constrains (\ref{trans2}).
The relevant physics of the Schwarzschild spacetime is well studied in the literature.

\

\noindent
\underline{Case II: $V(r)=\mathrm{const}=c_1$:}\\
Using the constrains $V(r)=\mathrm{const}=c_1$ in Eqs.~(\ref{fe}) and (\ref{chr}) we get
\begin{align}
\label{sol2}
h(r) = & \e^{\int^r \left( \frac{2\left(c_1 r'^3+3M \right)}{r \left( 6 r' +c_1 r7^3-6M \right)}
+ \frac{6 r'^{3/2}}{r'\left( 6 r'+c_1 r'^3-6M \right)^{3/2} \int^{r'} \frac {3\sqrt {r''} dr''}
{ \left( 6 r'' +c_1 r''^3 - 6 M \right) ^{3/2}}} \right) dr'} \nonumber\\
\approx& 1+\Lambda r^2 -\frac {2M}{r}-\frac {2M}{\Lambda r^3}+\frac {6 M^2 \Lambda-1}{3\Lambda^2 r^4}\,, \nonumber\\
h_1(r)=&1+\frac{c_1}{6}{r}^{2}-\frac {M}{r}=1+\Lambda{r}^{2}-\frac {M}{r}\,, \nonumber\\
\lambda(r)=& \frac{1}{\left( c_2-\bigintsss \!{\frac {3\sqrt {r}}{ \left( 6\,r+c_1{r}^{3}-6M\right) ^{3/2}}}{dr}\right)\sqrt {6
\,r+c_1{r}^{3}-6\,M}{r}^{3/2}}\approx {\frac {-36M}{{r}^{3}}}+\frac {M}{{\Lambda
}{r}^{5}}+\frac{2268M}{\Lambda{r}^{6}}\,, \nonumber\\
\zeta(r)=& \int{\frac{dr}{\sqrt{1+\Lambda{r}^{2}-\frac {M}{r}}}}\approx
{\frac {\ln\,r}{\sqrt {\Lambda}}}+\frac {1}{4{
\Lambda}^{3/2}{r}^{2}}-\frac {M}{6{\Lambda}^{3/2}{r}^{3}}
 -\frac {3}{32{\Lambda}^{5/2}{r}^{4}}\,.
\end{align}
where $\Lambda=\frac{c_1}{6}$ and we choose the constants of the integration to give the asymptotic behavior after $\approx$ for $h_1$ when $r$ is large.
By the notation $\approx$, we show the behavior when $r$ is large enough.
As Eq.~(\ref{sol2}) shows that the constant $c_1$ cannot have a zero value,
which means that it cannot reduce to the first case.

Note that $h(r)$ can be rewritten as
\begin{align}
\label{hr}
h(r) = & h_1(r) \e^{\int^r \frac{2dr'}{r' \left(h_1 \left( r' \right) \right)^{3/2} \int^{r'} \frac {dr''}{r'' \left(h_1 \left( r'' \right) \right)^{3/2}}} } \, .
\end{align}
For the solution $h_1(r)=0$
\begin{enumerate}
\item When $c_1>0$, only one real and positive solution.
\item When $c_1<0$,
\begin{enumerate}
\item If $\frac{2}{3}\sqrt{-\frac{2}{c_1}}>M$ or $0>M^2 c_1 > - \frac{8}{9}$,  there are two horizons,
the horizon with larger radius is the cosmological horizon and the horizon with smaller radius is the
BH horizon.
\item If $\frac{2}{3}\sqrt{-\frac{2}{c_1}}<M$ or $M^2 c_1 < - \frac{8}{9}$, there is no horizon.
\end{enumerate}
\end{enumerate}
Let  a solution of $h_1(r)=0$ be $r_0$ and we now investigate the behavior of $h(r)$ when $r\sim r_0$.
Then $h_1(r)$ can be approximated as $h_1(r) \sim h_1' \left( r_0 \right) \left( r - r_0 \right)$ and $h(r)$ can be approximated as,
\begin{align}
\label{hr2}
h(r) \sim & h_1' \left( r_0 \right) \left( r - r_0 \right) \e^{\int^r \frac{2dr'}{\left(  r' - r_0 \right)^{3/2} \int^{r'} \frac {dr''}{\left( r'' - r_0 \right)^{3/2}}} }
\sim C h_1' \left( r_0 \right) \left( r - r_0 \right) \e^{- \ln \left(  r' - r_0 \right)}
=  C h_1' \left( r_0 \right) \neq 0 \, .
\end{align}
Here $C$ comes from the constant of integration.
Therefore $h(r)$ does not vanish and therefore there is no horizon, which tells that the spacetime given in (\ref{sol2}) is not black hole.
It might be, however, interesting that $h_1(r)$ has zero(s) and therefore the signature of $h_1(r)$ changes.
There appears the region of Euclidian signature where the signature of the diagonal
components of the metric is $(+,+,+,+)$ or the region with two times where the signature is $(+,+,-,-)$.

\

\noindent
\underline{Case III: $\lambda(r)\neq0$ and $V(r)\neq 0$}\\

There are many solutions of the field equations (\ref{fe}) and (\ref{chr}), however, we will discuses the physical solution only which has metric potential $h$ behaves as AdS/dS and also the mimetic potential asymptote a constant value as it should for any physical model. The solution that we will discus takes the form:
\begin{align}
\label{sol3}
h(r) =& 1+c_4r^2-\frac{2M}{r}-\frac{c_4}{r^3}\,, \nonumber\\
h_1(r) =&-\frac { \left( c_4{r}^{5}+{r}^{3}-2\,M{r}^{2}-c_4 \right) \left( -
16\,{r}^{3}+96\,M{r}^{2}+c_5\,r-144\,{M}^{2}r-80\,c_4 \right) }{4
 \left( 5\,c_4-2\,{r}^{3}+6\,M{r}^{2} \right) ^{2}} \,,
 \nonumber\\
\approx & 1+c_4r^2-\frac{c_4c_5}{16}-\frac{2 \cal M}{r}+\frac{c_6}{r^2}\,,\nonumber\\
V(r)=&-\frac{1}{2\left( 5\,c_4-2\,{r}^{3}
+6\,M{r}^{2} \right) ^{3}{r}^{2}}\Bigg[96\,c_4{r}^{11}-864\,{r}^{10}Mc_4+2592\,c_4{r}^{9}{M}^{2}-2\,c_5\,{r}^{9}c_4+18\,c_4{r}^{8}Mc_5-720\,{r}^{8}{c_4}^{2}\nonumber\\
&-2592\,
c_4{r}^{8}{M}^{3}+2\,c_5\,{r}^{7}+2880\,{c_5}^{2}M{r}^{7}-5040\,{M}
^{2}{r}^{6}{c_4}^{2}+35\,c_5\,{r}^{6}{c_4}^{2}-576\,{r}^{6}c+4-2\,c_5\,{r}^{6}M-2400\,{c_4}^{3}{r}^{5}\nonumber\\
&+2784\,c_4{r}^{5}M+17\,c_5\,{r}^{4}c_4
-5472\,{M}^{2}{r}^{4}c_4-1080\,{r}^{3}{c_4}^{2}-28\,c_5
\,{r}^{3}Mc_4+4032\,c_4{M}^{3}{r}^{3}+1320\,{c_4}^{2}M{r}^{2}-10\,c_5
\,r{c_4}^{2}\nonumber\\
&+1440\,r{M}^{2}{c_4}^{2}+900\,{c_4}^{3}\Bigg]
\nonumber\\
\approx & 6c_4-\frac{c_4c_5}{8r^2}\,,\nonumber\\
\lambda(r)=& -\frac {1}{
 \left( 5\,c-2\,{r}^{3}+6\,M{r}^{2} \right) ^{3}{r}^{2}}\Bigg[5\,c_5\,{r}^{6}{c_4}^{2}-5\,c_5\,r{c_4}^{2}+c_5\,{r}^{9}c_4+4\,c_5\,{r}^{4}c_4-2\,c_5\,{r}^{6}M+c_5\,{r}^{7}-10\,c_5\,{r}^{3}Mc_4
 +720\,{c_4}^{2}M{r}^{7}\nonumber\\
&-240\,{r}
^{8}{c_4}^{2}-720\,{M}^{2}{r}^{6}{c_4}^{2}-300\,{c_4}^{3}{r}^{5}+1200\,c_4{r}^
{5}M-240\,{r}^{6}c_4-2160\,{M}^{2}{r}^{4}c_4-60\,{r}^{3}{c_4}^{2}+1440\,c_4{M}
^{3}{r}^{3}\nonumber\\
&-120\,{c_4}^{2}M{r}^{2}+720\,r{M}^{2}{c_4}^{2}+300\,{c_4}^{3}\Bigg]
\nonumber\\
\approx & \frac{c_4c_5}{8r^2}+\frac{c_4(9Mc_5-240c_4)}{8r^3}+\frac{52c_4c_5M^2-720c_4{}^2M+c_5}{8r^4}\,.
\end{align}
where
\begin{align}
{\cal M}=10\,{c_4}^{2}-2\,M-\frac{3Mc_4c_5 }{8}\,, \qquad c_6=-\frac {27{M}^{2}c_4 c_5}{16}-\frac{c_5}{16}+ 45\,c_4{}^2M \,,
\end{align}
Equation~(\ref{sol3}) shows that when $c_4=c_5=0$, we return to the well-known Schwarzschild BH of GR.
Using Eq.~(\ref{sol3}) we get the mimetic field in the form
\begin{align}
\label{scal3}
\zeta(r)=&\int \!\displaystyle\frac {2}{\sqrt {-{\frac { \left( c_4{r}^{5}+{r}^{3}-2\,M{r}^
{2}-c_4 \right) \left( -16\,{r}^{3}+96\,M{r}^{2}+c_5\,r-144\,{M}
^{2}r-80\,c_4 \right) }{ \left( 5\,c_4-2\,{r}^{3}+6\,M{r}^{2} \right) ^{2}
}}}}{dr} \nonumber\\
\approx& \frac {\ln \left( r \right) }{\sqrt {c_4}}-\frac {c_5}{64\sqrt {c_4}{r}^{2}}\frac {1}{4{c_4}^{3/2}{r}^{2}}-
\frac {M}{3{c_4}^{3/2}{r}^{3}}+\frac {5\sqrt {c_4}}{3{r}^{3}}-\frac {c_5\,M}{16\sqrt {c_4}{r}^{3}}\,.
\end{align}
Using Eq.~(\ref{scal3}) we can write the radial coordinate in terms of the mimetic field $\zeta$ as:
\begin{equation}
\label{rx}
r=\e^{\frac{1}{2}W \left( \frac { \left(c_5 c_4-16 \right) {\e^{-2 \zeta \sqrt {c_4}}}}{32c_4}
 \right)} + \zeta \sqrt {c_4}\,.
\end{equation}
Here $W(x)$ is the Lambert function defined by $x=W(x)\e^{W(x)}$.
Using Eq.~(\ref{rx}) in the asymptotic form of the Lagrange multiplier and the mimetic potential given by Eq.~(\ref{sol3}) we get
\begin{align}\label{vx}
\lambda(\zeta)=&\frac{\,c_4c_5}{8\Upsilon^{2}}+ \frac{\left(9\,Mc_4c_5-240\,{c_4}^{2}
\right)}{8\Upsilon^{3}}+\frac{\left( 54\,{M}^{2}c_4c_5-1440\,{c_4}^{2}M+c_5\right)}{8\Upsilon^{4}}\,,\nonumber\\
V(\zeta)=& \frac{1 }{16\Upsilon^{5}}\left\{ 2\,c_5\,c_4 \Upsilon^{3}+ 6c_4\left( 3\,Mc_5-80\,{c_4} \right) \Upsilon^{2}
+ 2\left( c_5-18Mc_4[80{c_4}+3\,{M}c_5] \right) \Upsilon + 25\left(c_5 -576\,{M}^{2}\right) {c_4}^{2} \right. \nonumber\\
& \left. + 60\left( 9\,c_5\,{M}^{3}-8 \right) c_4+14\,Mc_5 \right\}\approx V_0+V_1\zeta+V_2 \zeta^2\,,
\end{align}
where $\Upsilon= {\e^{\frac{1}{2}\,{ W} \left(\frac { \left( c_5\,c-16 \right) {\e^{-2\, \zeta\,\sqrt {c_4}}}}{32c_4} \right) + \zeta \,\sqrt {c_4}}} $,
$V_0$, $V_1$ and $V_2$ are constants structured by the constants $c_4$, $c_5$, and $M$.
Using Eq.~(\ref{sol3}) in (\ref{met}), when $r$ is large enough, we get the line element in the form:
\begin{align}
\label{elm3}
ds^2\approx \left( 1+c_4r^2-\frac{2M}{r}-\frac{c_4}{r^3}\right)dt^2 -\frac{dr^2}{1+c_4r^2-\frac{c_4c_5}{16}-\frac{2 \cal M}{r}
+\frac{c_6}{r^2}}-r^2 \left( d\theta^2 + \sin^2\theta d\phi^2 \right)\,.
\end{align}
As we will see soon, the parameter $c_4$ must be negative in order to avoid the singularity.
Therefore Eq.~(\ref{elm3}) shows that our black hole asymptotes as dS.

Using Eq.~(\ref{met}), we get the curvature invariants as:
\begin{align}
\label{inv33}
R_{\mu \nu \rho \sigma} R^{\mu \nu \rho \sigma}
=&
\frac {1}{4h^{4}{r}^{4}}\left\{4\,{r}^{4} h''^{2} h^{2} h_1{}^{2}+4\,{r}^{4}hh_1h'  \left( h h'_1-h_1 h'
 \right)h'' +{r}^{4} h'^{4}h_1{}^{2}-2\,{r}^{4} h'^{3}h_1h'_1h \right. \nonumber\\
& \left. +{r}^{2} h^{2} \left(h'_1{}^{2}{r}^{2}+8\, h_1{}^{2} \right) h'^{2}
+8 h^{4} \left( h'_1{}^{2}{r}^{2}+2\, \left(h_1-1  \right)
^{2} \right) \right\}\,, \nonumber\\
R_{\mu \nu } R^{\mu \nu }=&\frac {1}{8h^{4}{r}^{4}}\left.\{4\,{r}^{4} h''^{2}h^{2} h_1{}^{2}+4\,h  \left[h  \left( rh'_1
 +2\,h_1  \right)h'-r h'^{2}h_1 +2\, h^{2}h'_1  \right]{r}^{3}h_1 h'' +{r}^{4}h'^{4}h_1{}^{2} \right. \nonumber\\
&+{r}^{2}h^{2} \left( 12\,h_1{}^{2}+h'_1 {}^{2}{r}^{2} \right)h'^{2} -2\,{r}^{3}h h_1  \left( rh'_1 +2 \,h_1  \right) h'^{3} \nonumber\\
& \left. +4\,r h^{3} \left( 2\,h'_1rh_1 -4\,h_1 +4\, h_1{}^{2}+ h'_1{}^{2}{r}^{2} \right) h'
+4\,h^{4} \left(3  h'_1{}^{2}{r}^{2}+4r \left( h_1-1  \right)h'_1
+4\left(h_1 -1 \right)^{2} \right) \right\}\,,\nonumber\\
R=&\frac {2\,h'' h_1 h  {r}^{2}- h'^{2}h_1 {r}^{2}+rh   \left( 4\,h_1 +rh'_1  \right) h'  +4\,h^{2} \left( rh'_1+h_1-1  \right) }{2{r}^{2}h^{2}}\,.
\end{align}
Here $R_{\mu \nu \rho \sigma} R^{\mu \nu \rho \sigma}$,
$R_{\mu \nu} R^{\mu \nu}$, and $R$ represent
the Kretschmann scalar, the Ricci tensor square, and the Ricci scalar, respectively.

For the solution in (\ref{sol3}), when $r$ is large, we find
\begin{align}
\label{inv22}
R_{\mu \nu \rho \sigma} R^{\mu \nu \rho \sigma}\approx &24c_4{}^2-\frac {3c_4{}^2c_5}{2{r}^{2}}+\frac {3c_4{}^2(160c_4+3c_5M)}{4{r}^{3}}
+\frac {c_4{}M(3c_4M+32)}{64{r}^{4}}\,, \nonumber\\
R_{\mu \nu } R^{\mu \nu }\approx & 36c_4{}^2-\frac {9c_4{}^2c_5}{4{r}^{2}}
+\frac {9c_4{}^2(160c_4+3c_5M)}{8{r}^{3}}+\frac {3c_4{}M(16+c_4M)}{64{r}^{4}}\,, \nonumber\\
R\approx & 12c_4-\frac {3c_4c_5}{8{r}^{2}}+\frac {3c_4(160c_4+3c_5M)}{16{r}^{3}}+\frac {c_5}{8{r}^{4}}\,.
\end{align}
On the other hand, when $r$ is small, we find,
\begin{align}
\label{inv22B}
R_{\mu \nu \rho \sigma} R^{\mu \nu \rho \sigma} \sim R_{\mu \nu } R^{\mu \nu } \sim R \sim \mathcal{O}\left(r^{-3}\right)\,.
\end{align}
Equation (\ref{inv22B}) shows that the BH solution (\ref{sol3}) has a softer singularity at $r=0$ compared with the BH in general relativity,
where $\left( R_{\mu \nu \rho \sigma}
R^{\mu \nu \rho \sigma}, R_{\mu \nu} R^{\mu \nu},
R \right) = \left( \frac{24(\Lambda^2 r^6+2M^2)}{r^6}, 36\Lambda^2, 12\Lambda \right)$.

The curvature invariants in (\ref{inv33}) might diverge when $h(r)=0$ because the expressions in (\ref{inv33}) include the inverse
power of $h(r)$.
We should note, however, $h_1(r)$ given by Eq.~(\ref{sol3}) can be rewritten as follows,
\begin{align}
\label{h1A1}
h_1(r) =\frac{r^3 h(r) \left( r \left( 4r - 12 M \right)^2 - c_5 r + 80 c_4 \right)}{4 \left( 5 c_4 - 2r^3 + 6 M r^2 \right)^2} \, .
\end{align}
Therefore when $h(r)$ vanishes, $h_1(r)$ also vanishes.
This tells that the zeros of  $h_1(r)$ cancels the zeros of $h(r)$ and the curvature invariants (\ref{inv33}) become finite and do
not diverge.
We now show that in this case, the invariants in (\ref{inv33}) do not diverge but are finite at the zeros $h(r)=0$ of $h(r)$.
We consider the case that $h_1$ is written by $h_1 = h_2 h$ as in (\ref{h1A1}).
Here $h_2$ does not vanish and is regular at the zeros $h(r)=0$ of $h(r)$.
By substitute the expression $h_1 = h_2 h$ into (\ref{inv33}), we find
\begin{align}
\label{invariants2}
R_{\mu \nu \rho \sigma} R^{\mu \nu \rho \sigma}
=& {h''}^2 h_2^2 + h_2 h_2' h' + \frac{{h'}^2 {h_2'}^2}{4} + \frac{4h_2^2 {h'}^2}{r^2} + \frac{2\left( \left( h_2' h + h_2 h' \right)^2 r^2 + 2 \left( h_2^2 h^2 - 1 \right)^2 \right)}{r^4} \, , \\
\label{invariants3}
R_{\mu \nu } R^{\mu \nu }
=& \frac{{h''}^2 h_2^2}{2} + \frac{h'' h' h_2 h_2'}{2} + \frac{h' h'' h_2^2}{r} + \frac{\left(h_2' h + h_2 h'\right) h^2 h_2}{r}
+ \frac{3{h'}^2 h_2^2}{2r^2} + \frac{{h'}^2 {h_2}^2 }{8} + \frac{h' h_2 \left( h_2' h + h_2 h' \right)}{r^2} \nn
& - \frac{2h' h_2^2}{r^3} + \frac{2h h' h_2^2}{r^3}
+ \frac{h h' h_2^2}{2r} + \frac{{h'}^2 h_2 h_2'}{r} \frac{3 \left( h_2' h + h_2 h' \right) r^2
+4r \left( h_2 h-1  \right) \left( h_2' h + h_2 h' \right) +4\left( h_2 h -1 \right)^2}{2r^4} \, , \\
\label{invariants1}
R=&2h'' h_2 + \frac{2r h_2 h'}{r} + \frac{h_2' h'}{2} + \frac{2\left( r h_2' h + r h_2 h' + h_1 - 1\right)}{r^2} \, .
\end{align}
Therefore the invariants are surely finite at the zeros $h(r)=0$.

We should also note that the denominator in the expression (\ref{h1A1}) can vanish in general. If the denominator vanishes, $h_1(r)$
diverges and therefore the curvature invariants (\ref{inv33}) diverge and a singularity appears.
The singularity can be avoided if we choose the parameter $c_4$ satisfy the following condition,
\begin{align}
\label{Consh1}
5 c_4 + 8 M^3 <0 \, ,
\end{align}
because the maximum of $- 2r^3 + 6 M r^2 $ is $8 M^3$ when $r=2M$.

It might be also interesting that the factor $r \left( 4r - 12 M \right)^2 - c_5 r + 80 c_4$ can vanish and
it generates an extra zero(s) besides the zeros of $h(r)$.
The existence of the zero tells that there appears the region of Euclidian signature where the signature of the diagonal
components of the metric is $(+,+,+,+)$ or the region with two times where the signature is $(+,+,-,-)$.
Such regions might be interesting but could not be physically acceptable.
Let a radius of the black hole horizon(s) be $r_\mathrm{BH}$.
If we choose the parameter $c_5$ to satisfy
\begin{align}
\label{Consh1B}
 - c_5 r_\mathrm{BH} + 80 c_4 > 0 \, ,
\end{align}
as long as $r>r_\mathrm{BH}$, the factor $r \left( 4r - 12 M \right)^2 - c_5 r + 80 c_4$ is positive and does not vanish and therefore
the unphysical region(s) is hidden inside the black hole horizon and the existence of the region(s) becomes less unphysical.

\begin{figure}[ht]
\centering
\includegraphics[scale=0.35]{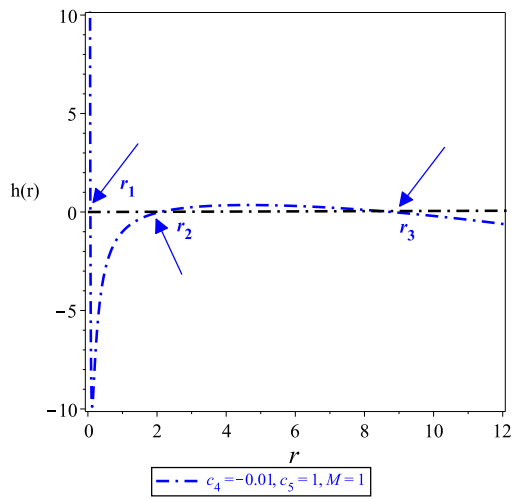}\hspace{0.2cm}
\caption{ {Multi-horizon plot of Eq.~(\ref{hor333}) against coordinate $r$ for the BH (\ref{sol3}).}}
\label{Fig:1}
\end{figure}

Because the equation $h(r)=0$ has three real positive solutions, we may write $h(r)$ in the following form,
\begin{equation}
\label{hor333}
h(r)=1+c_4r^2-\frac{2M}{r}-\frac{c_4}{r^3}=h_3(r)(r-r_1)(r-r_2)(r-r_3)\,,
\end{equation}
where we assume $r_1<r_2<r_3$.
We plot Eq.~(\ref{hor333}) for specific values of the mass, the constants $c_4$, and $c_5$ in Figure~\ref{Fig:1}.

\begin{figure*}
\centering
\subfigure[~The lagrangian multiplier for $M=0.1$, $c_5=100$]{\label{fig:2a}\includegraphics[scale=0.27]{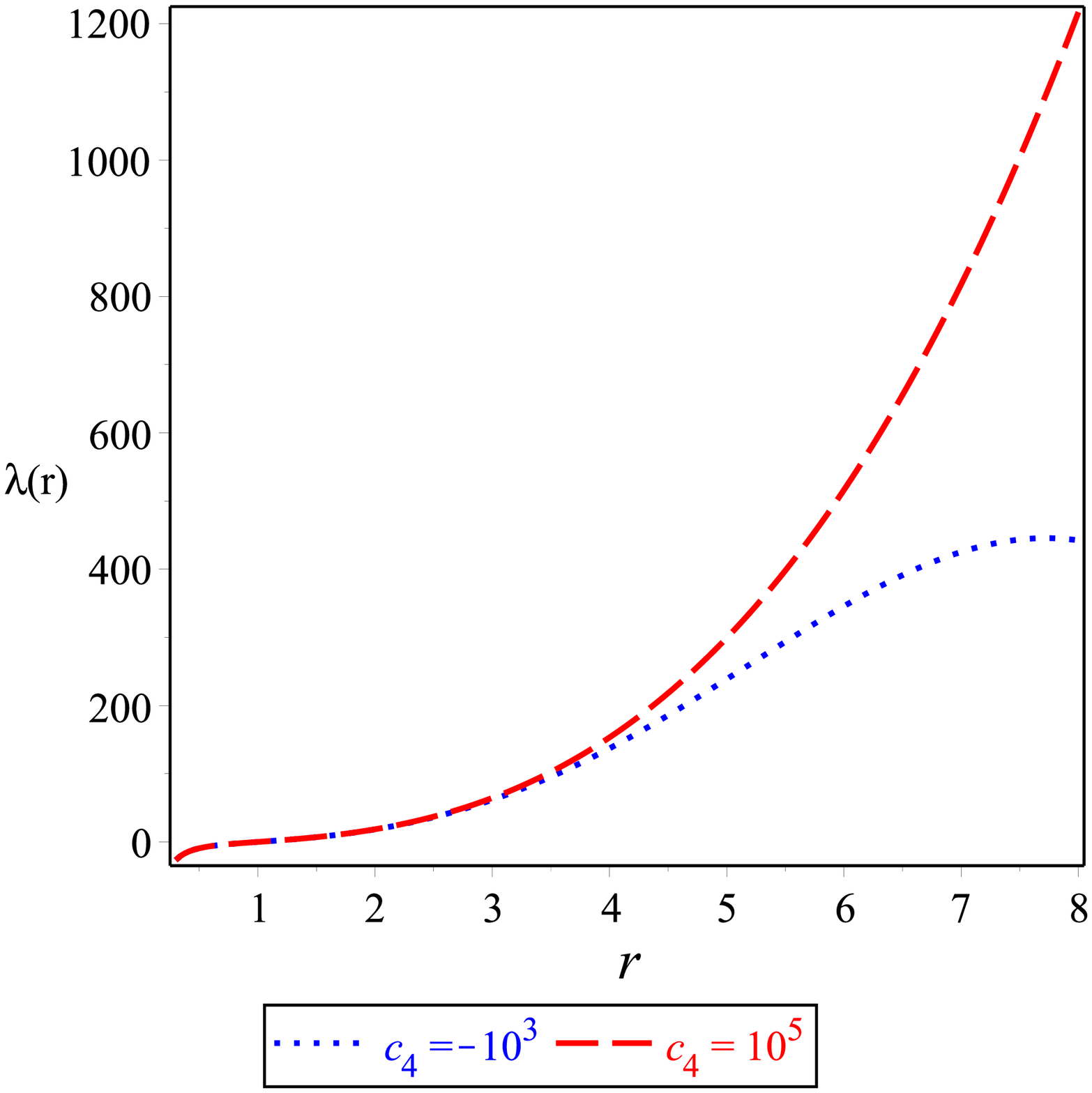}}\hspace{0.5cm}
\subfigure[~The mimetic potential for $M=0.1$]{\label{fig:2b}\includegraphics[scale=0.27]{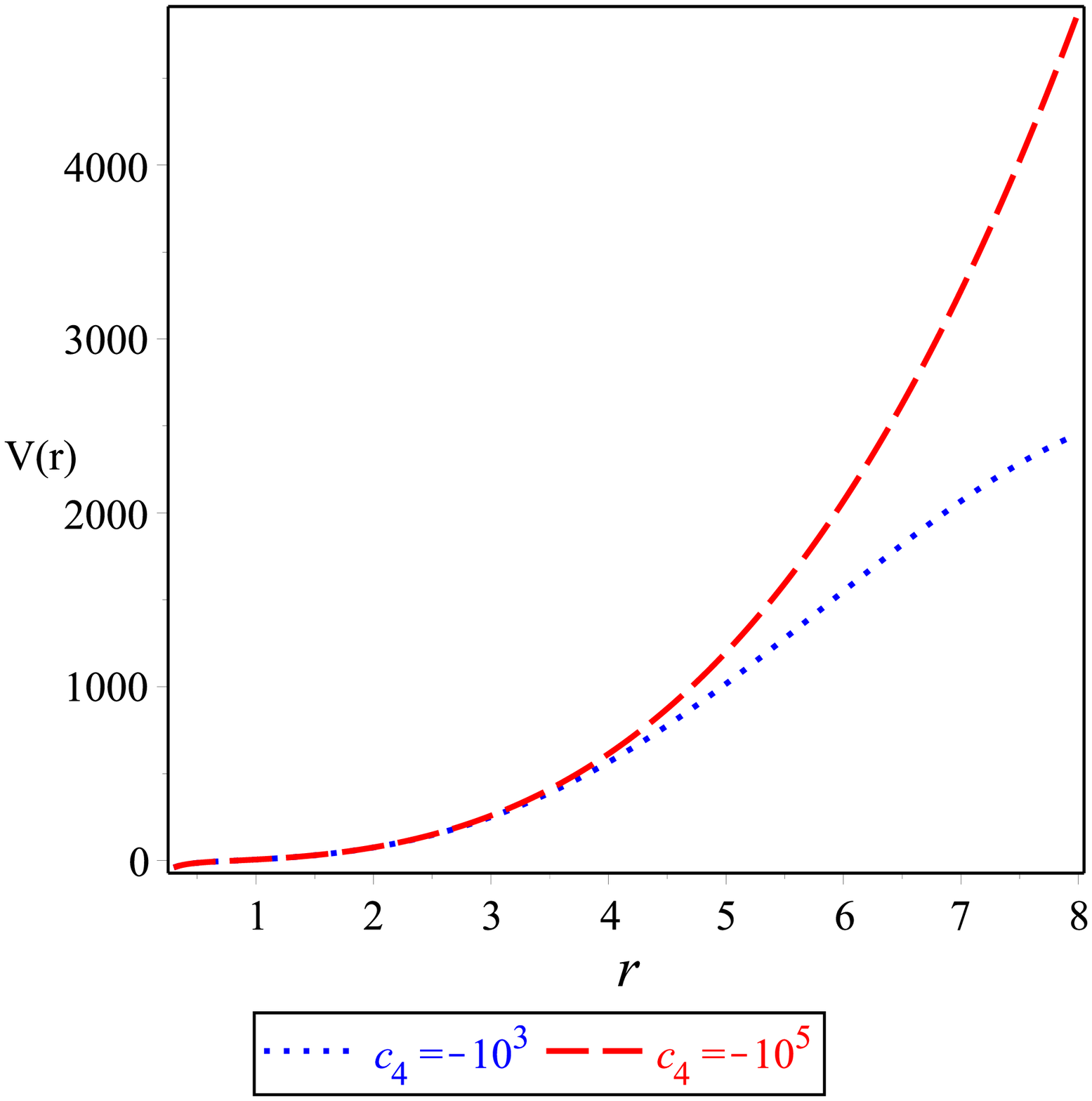}}
\subfigure[~The mimetic field for $M=0.1$, $\Lambda=100$]{\label{fig:2c}\includegraphics[scale=0.27]{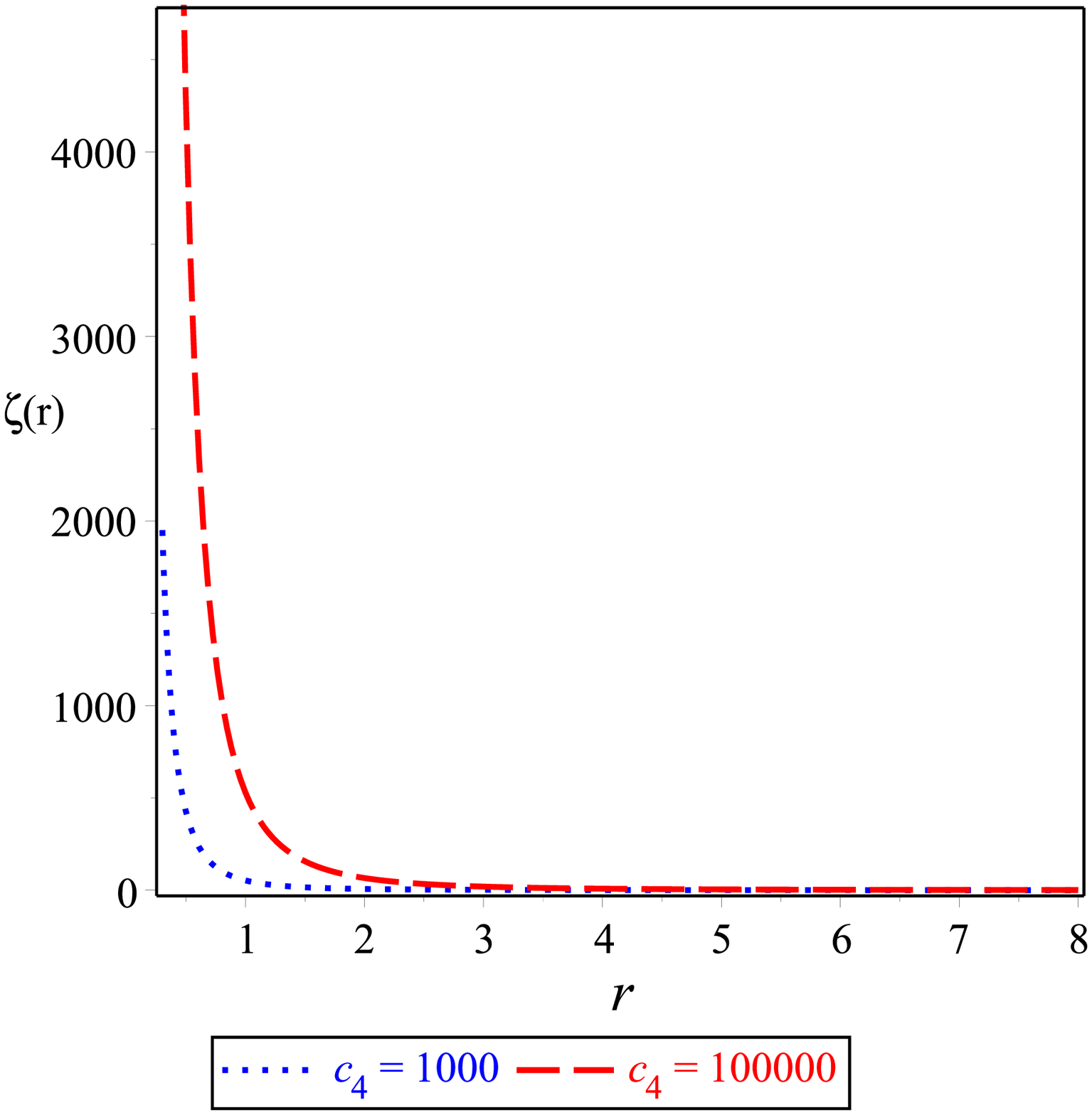}}\hspace{0.5cm}
\caption{~\subref{fig:2a} Behavior of the Lagrange multiplier of Eq.~(\ref{sol3}); ~\subref{fig:2b}the mimetic potential of Eq.~(\ref{sol3}); and \subref{fig:2c} the mimetic field of Eq.~(\ref{sol3}).}
\label{Fig:2}
\end{figure*}

The behavior of the Lagrange multiplier, the  mimetic potential, and the mimetic field are drawn in Figure~\ref{Fig:2}~\subref{fig:2a},  Figure~\ref{Fig:2}~\subref{fig:2b}
and Figure~\ref{Fig:2}~\subref{fig:2c}, where they   have positive value.

\section{Energy conditions}\label{S3}

Energy conditions provide important tools to examine and better understand cosmological models and/or strong gravitational fields.
We are interested in the study of energy conditions in the mimetic theory coupled with Lagrange multiplier and mimetic potential,
because this is the first time to derive a non-trivial BH solutions in Case III given by (\ref{sol3}).
The energy conditions are classified into four categories: strong energy (SEC), weak energy (WEC), null energy (NEC),
and dominant energy (DEC) conditions \cite{Hs73, Nashed:2016gwp}.
To fulfill these conditions, the following inequalities should be verified:
\begin{align}
\label{econ}
\mbox{SEC}: & \quad \rho+p_r\geq 0\,, \quad \rho+p_t\geq 0\,, \quad \rho-p_r-2p_t\geq 0\, , \nonumber\\
\mbox{WEC}: & \quad \rho\geq 0\,, \quad \rho+p_r\geq 0\,, \nonumber\\
\mbox{NEC}: & \quad \rho \geq 0\, , \quad \rho+p_t\geq 0\, , \nonumber\\
\mbox{DEC}: & \quad \rho\geq \left|p_r\right|\, , \quad \rho\geq \left| p_t \right|\, ,
\end{align}
where the energy-momentum tensor ${T^\zeta}_{\mu \nu}$ for the mimetic field $\zeta$ and
the Lagrange multiplier field $\lambda$ is defined as:
\begin{align}
\label{EM}
{T^{\zeta}}_{\mu \nu}=\frac{1}{2}g_{\mu \nu}V(\zeta)+\lambda
\partial_{\mu}\zeta \partial_{\nu}\zeta\,,
\end{align}
with ${T^{\zeta}}_0{}^0=\rho$, \, ${T^{\zeta}}_1{}^1=-p_r$ and ${T^{\zeta}}_2{}^2={T^{\zeta}}_3{}^3=-p_t$
are the density, radial, and tangential pressures, respectively.
Note that the constraint (\ref{trans2}) is used in Eq.~(\ref{EM}).
The form of the components of the energy-momentum tensor have the form:
\begin{align}
\label{em1}
\rho=\frac{V}{2}=\frac{c_1}{2}\,, \quad p_r=\frac{V}{2}-\lambda(r)h_1(r)\zeta'\, , \quad p_t=\frac{V}{2}=\frac{c_1}{2}\, .
\end{align}
Straightforward calculations of the BH solution (\ref{sol3}) give
\begin{align}
\label{econ1}
\mbox{SEC}: \quad & \rho+p_r = \frac { \left(c_5\,{r}^{4}-240\,c_4{r}^{3}+720\,{r}^{2}Mc_4+5\,c_5r
-300\,{c_4}^{2} \right) \left(c_4 -c_4{r}^{5}-{r}^{3}+2\,M{r}^{2} \right)}{ \left( 5\,c_4-2\,{r}^{3}+6\,M{r}^{2} \right) ^{3}{r}^{2}}
> 0\, ,\quad \rho+p_t=0\, , \nonumber\\
& \rho-p_r-2p_t = \frac {1}{ \left( 5\,c_4-2\,{r}^{3} + 6\,M{r}^{2} \right) ^{3}{r}^{2}} \left[96r^{10}c_4[9M-{r}]+ 3\left(c_5-864 {M}
^{2} \right) c_4{r}^{9} \right.\nonumber\\
& \qquad + 6\left[80\,{c_4}^{2}+ 3M\left( 144\,{M}^2-c_5\right) c_4 \right] {r}^{8}
 -\left( 2160\,{c_4}^{2}M+c_5 \right) {r}^{7}+ 3c_4\left[10 \left(144\,{M}^{2}-c_5
\right) {c_4}+112 \right] {r}^{6} \nonumber \\
& \qquad + 12c_4\left( 175\,{c_4}^{2}-132\,M
\right) {r}^{5}+c_4 \left( 3312\,{M}^{2}-13\,c_5 \right) {r}^{4}\nonumber\\
& \left. \qquad + 6\left[ 170\,{c_4}^{2}+ 3\left( c_5\,M-144\,{M}^{3}
 \right) c_4 \right]{r}^{3}-1440\,{c_4}^{2}M{r}^{2}- 5\left( 144\,{M}^{2}
-c_5 \right) {c_4}^{2}r-600\,{c_4}^{3} \right]
> 0\, , \nonumber\\
\mbox{WEC}: \quad & \rho= \frac{1}{4 \left( 5\,c_4-2\,{r}^{3}+6\,M{r}^{2} \right) ^{3}{r}^{2}} \left[ 96{r}^{10}c_4(9\,M -1)+ \left( 2\,c_5-2592
\,{M}^{2} \right) c_4{r}^{9}+ \left[ 720\,{c_4}^{2}+ 18M\left(144\,{M}^{2} -c_5 \right) c_4 \right] {r}^{8} \right. \nonumber\\
& \qquad - 2\left( 1440\,{c_4}^{2}M+\,c_5 \right) {r}^{7}+ 35\left[ \left( 144\,{M}^{2}-c_5 \right) {c_4}^{2}+576\,c_4+2\,c_5\,M \right] {r}^{6} \nonumber \\
& \qquad + \left( 2400\,{c_4}^{3}-2784\,c_4M \right) {r}^{5}+c_4 \left( 5472\,{M}^{2}-17\,c_5 \right) {r}^{4}\nonumber\\
& \left. \qquad + \left[ 1080\,{c_4}^{2}+ 28\left( c_5
\,M-144\,{M}^{3} \right) c_4 \right] {r}^{3}-1320\,{c_4}^{2}M{r}^{2}+10
\, \left( c_5-144{M}^{2} \right) {c_4}^{2}r-900\,
{c_4}^{3} \right] > 0\, , \nonumber\\
& \rho+p_r = \frac { \left(c_5\,{r}^{4}-240\,c_4{r}^{3}+720\,{r}^{2}Mc_4+5\,c_5r
 -300\,{c_4}^{2} \right) \left(c_4 -c_4{r}^{5}-{r}^{3}+2\,M{r}^{2} \right)
}{ \left( 5\,c_4-2\,{r}^{3}+6\,M{r}^{2} \right) ^{3}{r}^{2}}
> 0\,, \nonumber\\
\mbox{NEC}: \quad & \rho=\frac{1}{4 \left( 5\,c_4-2\,{r}^{3}+6\,M{r}^{2} \right) ^{3}{r}^{2}}\left[96{r}^{10}c_4(9\,M -1)+ \left( 2\,c_5-2592
\,{M}^{2} \right) c_4{r}^{9}+ \left[ 720\,{c_4}^{2}+ 18M\left(144\,{M}^{2} -c_5 \right) c_4 \right] {r}^{8} \right. \nonumber\\
& \qquad - 2\left( 1440\,{c_4}^{2}M+\,c_5 \right) {r}^{7}+ 35\left[ \left( 144\,{M}^{2}-c_5 \right) {c_4}^{2}+576\,c_4+2\,c_5\,M \right] {r}^{6} \nonumber \\
& \qquad + \left( 2400\,{c_4}^{3}-2784\,c_4M \right) {r}^{5}+c_4 \left( 5472\,{M}^{2}-17\,c_5 \right) {r}^{4}\nonumber\\
& \left. \qquad + \left[ 1080\,{c_4}^{2}+ 28\left( c_5
\,M-144\,{M}^{3} \right) c_4 \right] {r}^{3}-1320\,{c_4}^{2}M{r}^{2}+10
\, \left( c_5-144{M}^{2} \right) {c_4}^{2}r-900\,
{c_4}^{3} \right]> 0\, , \nonumber\\
& \rho+p_t=0\, , \nonumber\\
\mbox{DEC}: \quad & \rho\geq \left|p_r\right| \quad \mbox{(satisfied)}\, , \quad \rho\geq \left| p_t \right| \quad \mbox{(satisfied)}\, .
\end{align}

\begin{figure*}
\centering
\subfigure[~WEC]{\label{fig:3a}\includegraphics[scale=0.2]{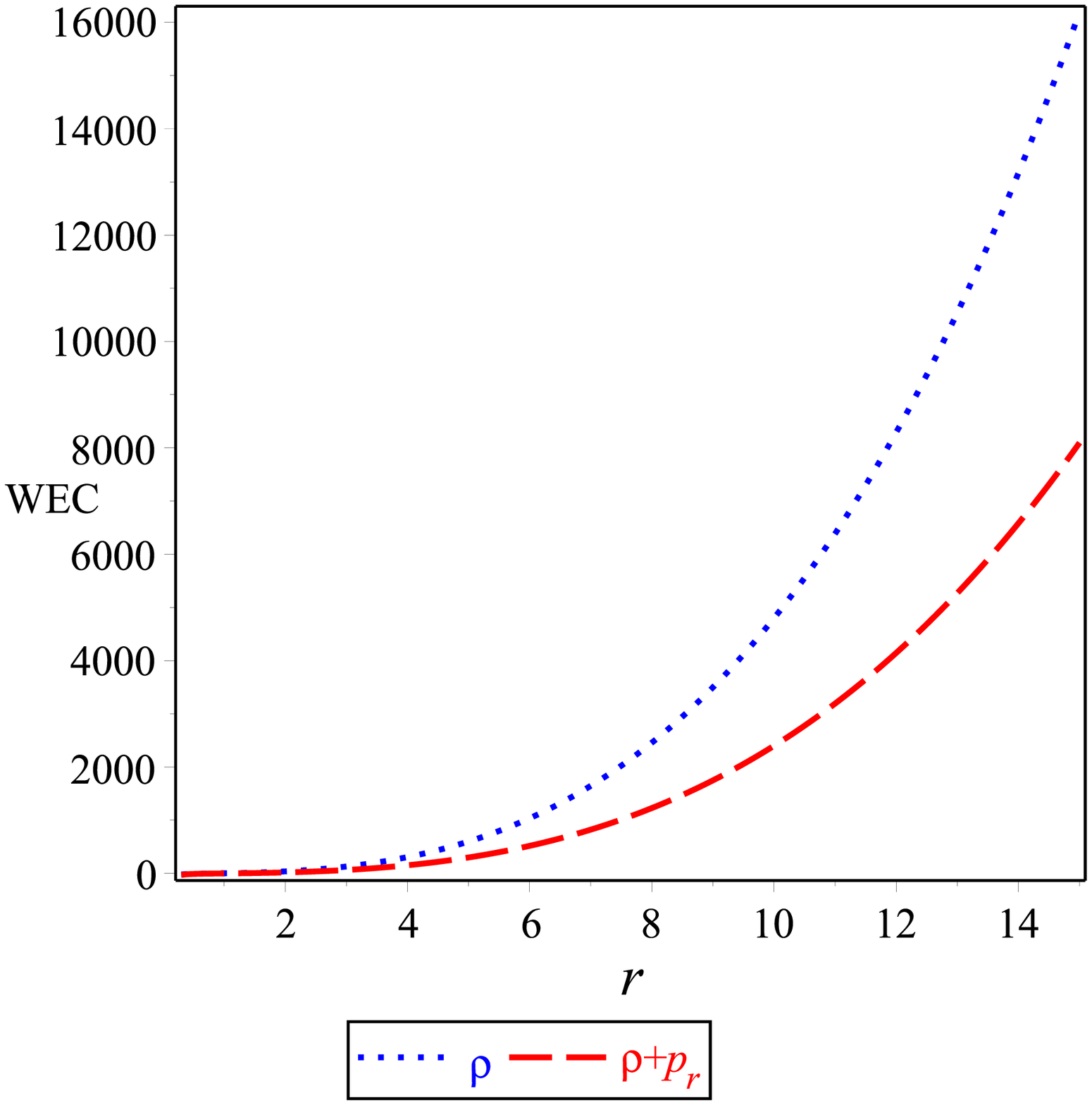}}\hspace{0.5cm}
\subfigure[~SEC]{\label{fig:3b}\includegraphics[scale=0.2]{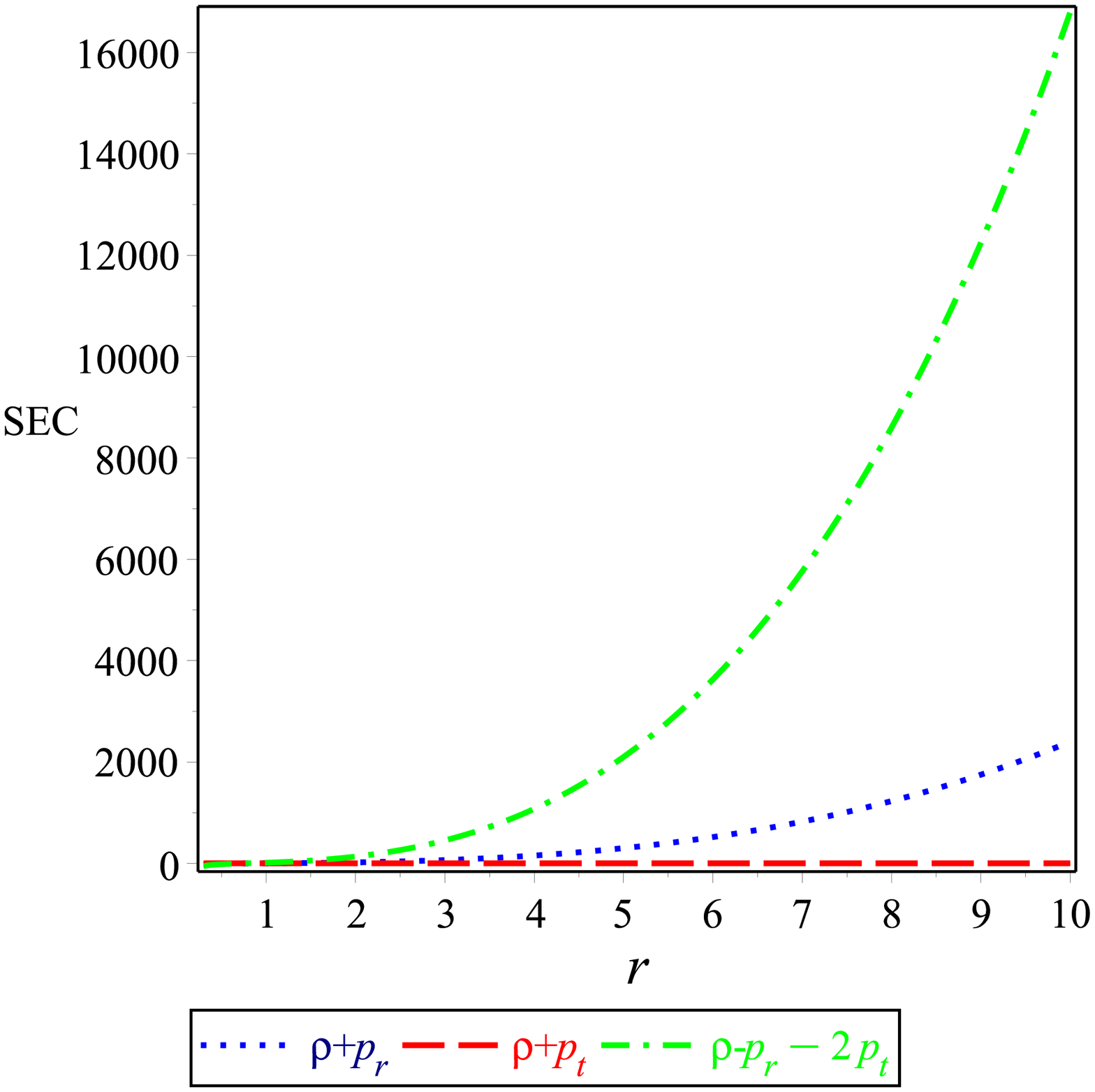}}\hspace{0.5cm}
\subfigure[~NEC]{\label{fig:3c}\includegraphics[scale=0.2]{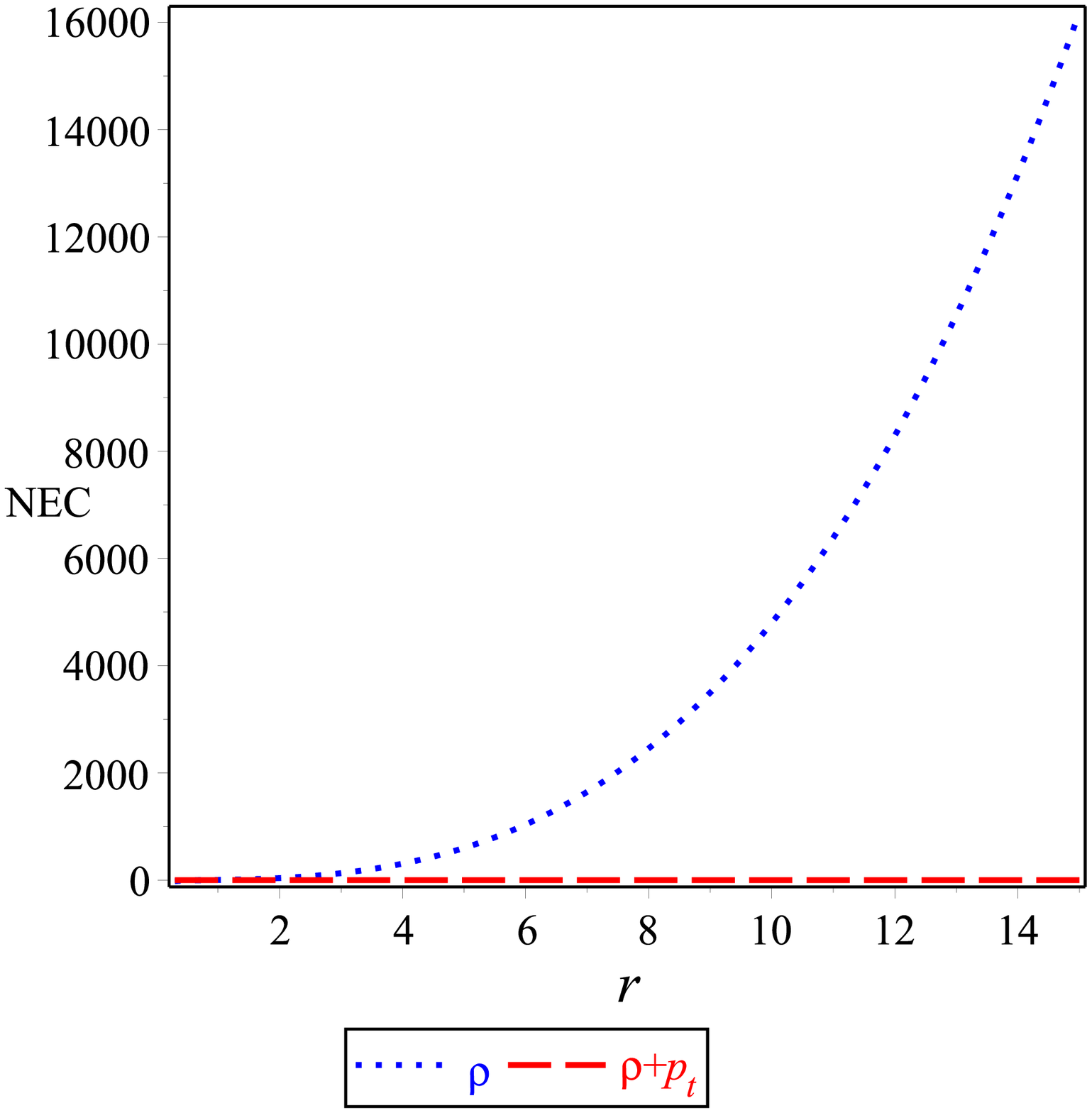}}\hspace{0.5cm}
\subfigure[~DEC]{\label{fig:3d}\includegraphics[scale=0.2]{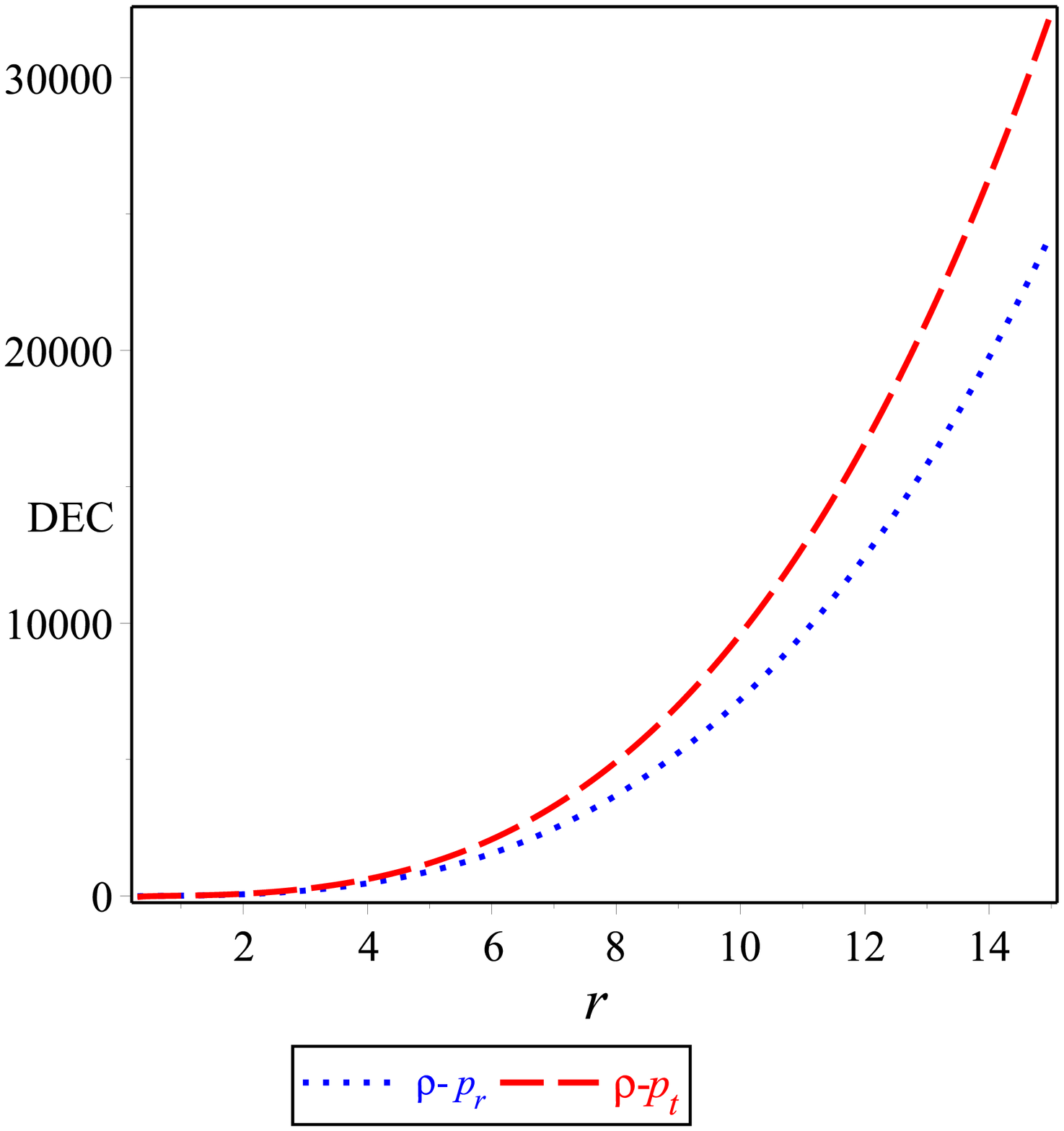}}
\caption{Schematic plots of WEC, SEC, NEC and DEC given by Eq.~(\ref{sol3}). All the above figures are plotted for $M=0.93$, $c_4=-10^{7}$ and $c_5=100$\footnote{We stress that the values of $M$ and $c_4$ taken in Figures~\ref{Fig:2}, Figures~\ref{Fig:3}  and Figures~\ref{Fig:4} did not make the value of $h1(r)$ given by Eq.~(\ref{sol3}) to have any imaginary value,  i.e., $c_4<-\frac{8M^3}{5}$ and $c_4<0$.}.}
\label{Fig:3}
\end{figure*}
Figures~\ref{Fig:3}~\subref{fig:3a}--\ref{Fig:3}~\subref{fig:3d} shows the energy conditions of the BH solution (\ref{sol3}), which also coincides with Eq.~(\ref{econ}).
The main reason that the energy conditions of the BH solution (\ref{sol3}) is the non-trivial form of the mimetic potential and its physical asymptote behavior.

\section{Thermodynamics and stability }\label{S4}
We consider another physics approach to deeply elucidate the two BHs with (\ref{sol3}) by investigating their thermodynamic behavior.
Accordingly, we will present the main tools of the thermodynamic quantities.

The entropy is given by \cite{PhysRevD.84.023515,Zheng:2018fyn}
\begin{equation}
\label{ent11}
S(r_+)=\frac{1}{4}A=\pi {r_+}^2\, ,
\end{equation}
where $A$ is the area of the horizon and the horizon radius $r_+$ satisfies the condition $h\left(r_+\right)=0$.

\begin{figure*}
\centering
\subfigure[~The horizon mass-radius]{\label{fig:4a}\includegraphics[scale=0.27]{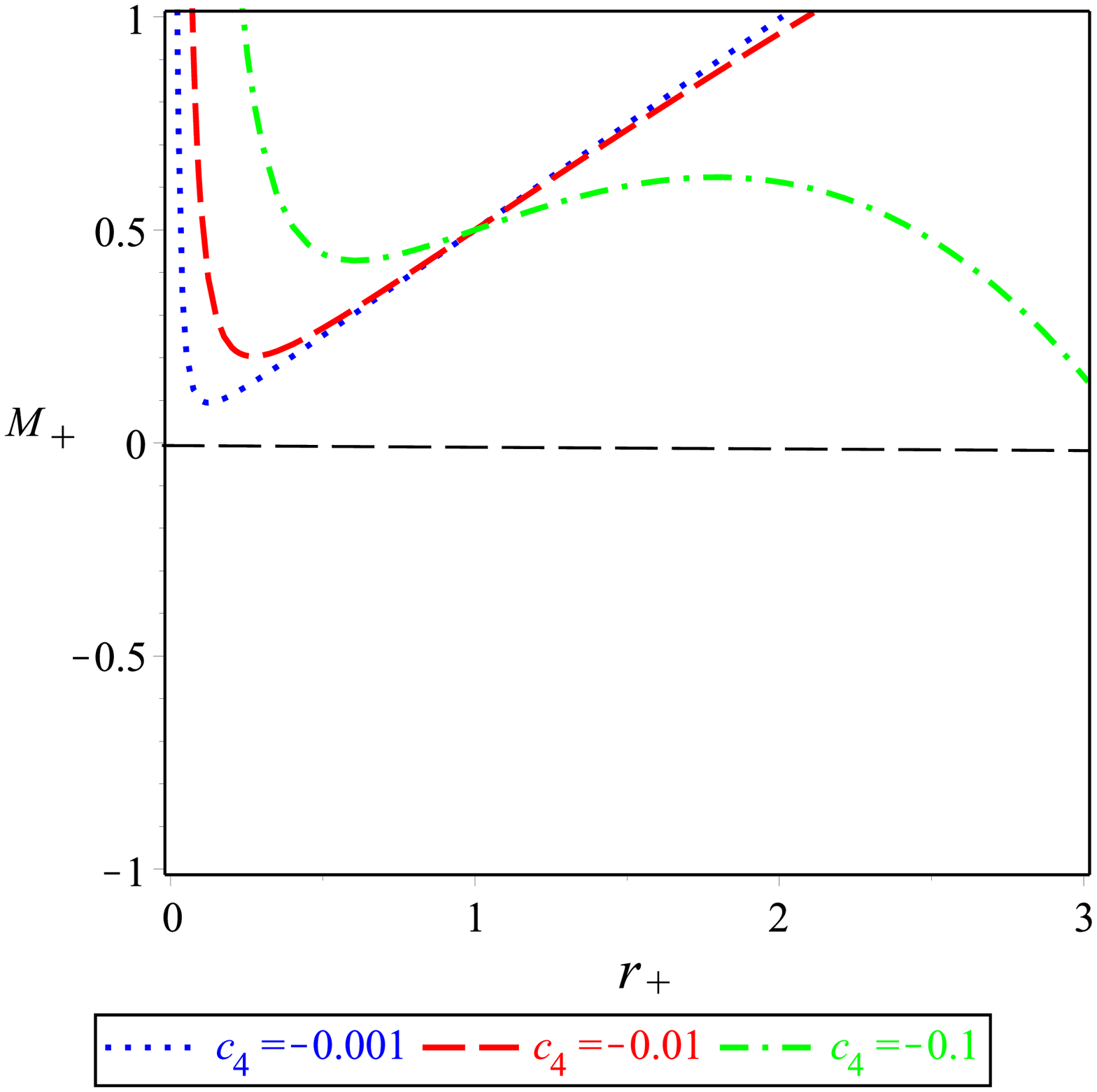}}\hspace{0.5cm}
\subfigure[~The horizon Hawking temperature for $c_4=-0.001$
]{\label{fig:4b}\includegraphics[scale=0.27]{
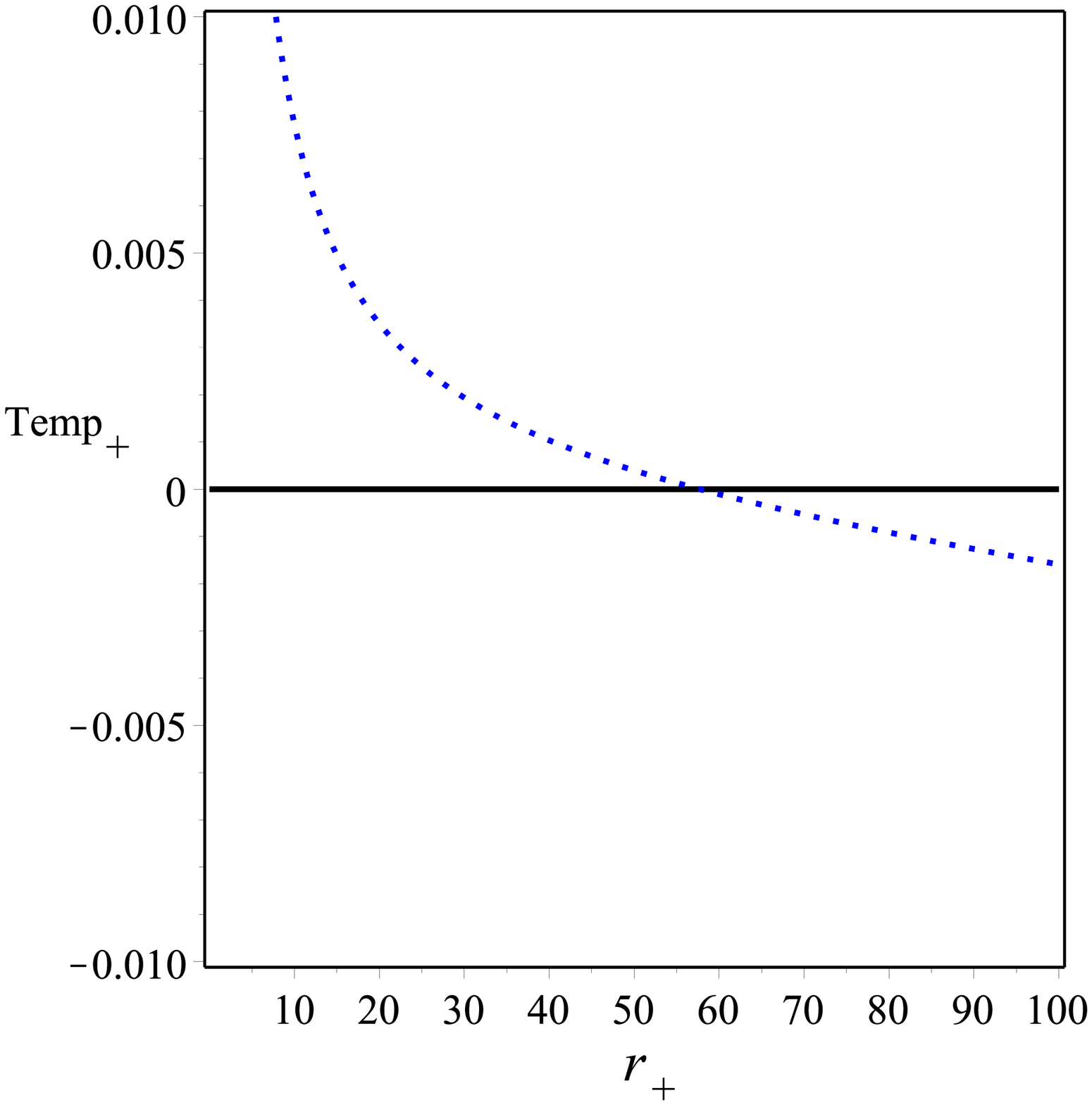}}\hspace{0.5cm}
\subfigure[~The horizon heat capacity for $c_4=-0.001$
]{\label{fig:4c}\includegraphics[scale=0.27]{
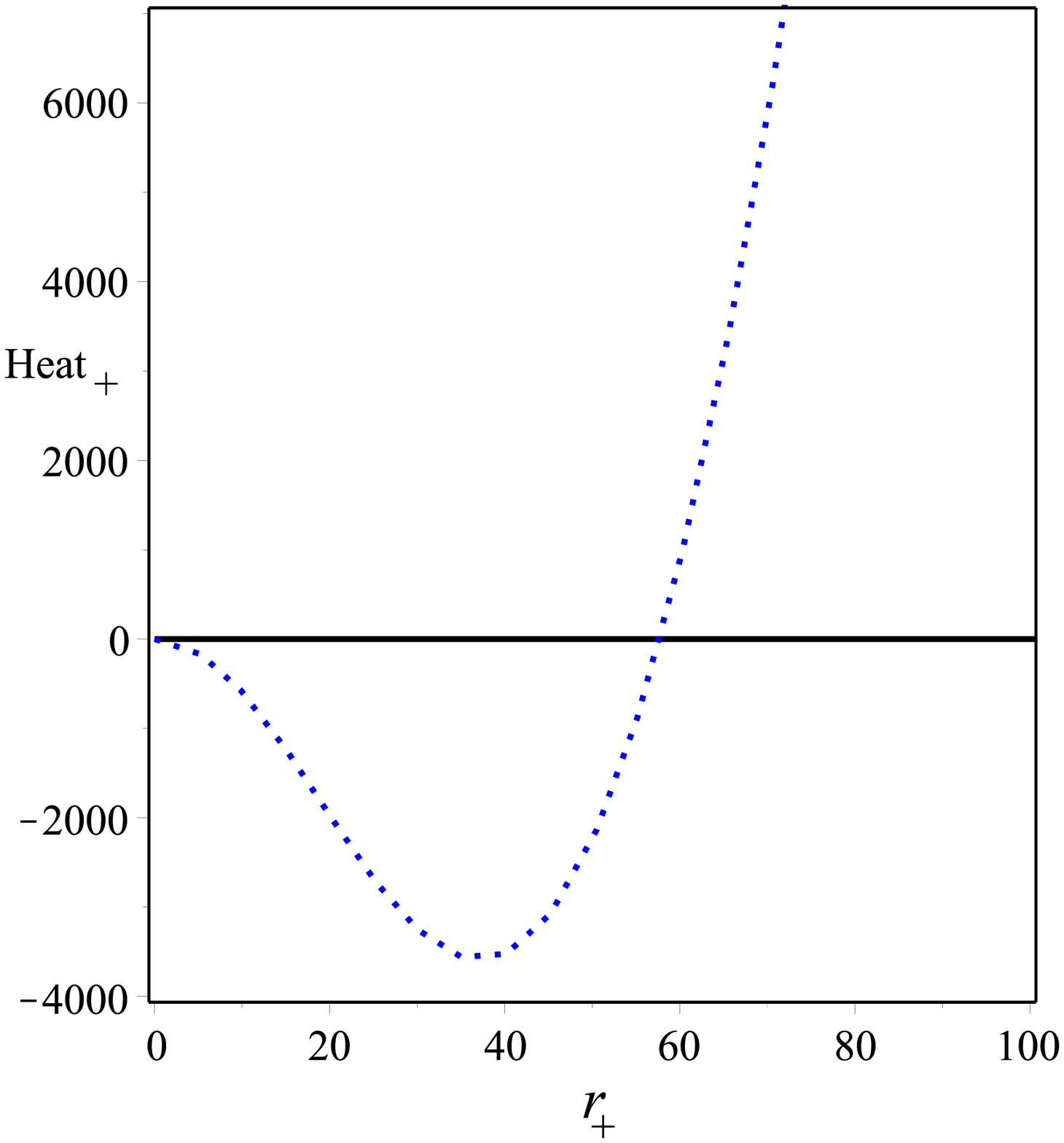}}\hspace{0.5cm}
\caption{Schematic plots of the thermodynamic quantities of the BH solution (\ref{sol3}) for a negative values of the constant $c_4$:~\subref{fig:4a},
the mass--radius relation of the horizon (\ref{hor-mass-rad1a33});~\subref{fig:4b},
typical behavior of the temperature of horizon (\ref{T22g})
for $c_4=-0.001$. There appears the critical point $r_+\sim 60$, which corresponds to the extremal limit.
And~\subref{fig:4c} the heat capacity, (\ref{heat-cap1a33}), which vanishes at the extremal limit $r_+\sim 60$.
indicates that we obtain a negative heat capacity
only when $r_+$ is smaller than the critical point $r_+\sim 60$.
}
\label{Fig:4}
\end{figure*}

The total mass contained within the event horizon ($r_+$) can be obtained by solving the equation $h(r_+) = 0$ with respect to $M$,
\begin{align}
\label{hor-mass-rad1a33}
M\left(r_+\right) =\frac {c_4{r_+}^{5}+{r_+}^{3}-c_4}{2{r_+}^{2}}\, .
\end{align}
The behavior of Eq.~(\ref{hor-mass-rad1a33}) is drawn in Figure~\ref{Fig:4}~\subref{fig:4a}.

The Hawking temperature is generally defined as follows \cite{PhysRevD.86.024013,Sheykhi:2010zz,Hendi:2010gq,PhysRevD.81.084040}
\begin{equation}
\label{temp11}
T_+ = \frac{h'(r_+)}{4\pi}\, ,
\end{equation}
where we may assume $h'(r_+)\neq 0$.
Using Eq.~(\ref{temp11}), the Hawking temperature of BH (\ref{sol3}) can be calculated as:
\begin{align}
\label{T22g}
T_+ = \frac {2c_4{r_+}^{5}+2\,M\left(r_+\right) {r_+}^{2}+3\,c_4}{4\pi \,{r_+}^{4}}
=\frac {3 c_4{r_+}^{5} + {r_+}^3 + 2 c_4}{4\pi \,{r_+}^{4}} \,,
\end{align}
The behavior of the Hawking temperature given by Eq.~(\ref{T22g}) is displayed in Figure~\ref{Fig:4}~\subref{fig:4b}.
There appears the critical point, which corresponds to the extremal limit where the radii of the two horizons coincide with each other.

The stability of the BH solution is an essential topic that can be studied at the dynamic and perturbative levels
\cite{Nashed:2003ee,Myung:2011we,Myung:2013oca}.
To investigate the thermodynamic stability of BHs, the formula of the heat capacity $H(r_+)$
at the event horizon must be derived. It is defined as follows \cite{Nouicer:2007pu,DK11,Chamblin:1999tk}:
\begin{equation}
\label{heat-capacity11}
H_+\equiv H(r_+)=\frac{\partial M_+}{\partial T_+}=\frac{\partial M_+}{\partial r_+} \left(\frac{\partial T_+}{\partial r_+}\right)^{-1}\, .
\end{equation}
The BH will be thermodynamically stable, if its heat capacity $H_+$ is positive.
On the other hand, it will be unstable if $H_+$ is negative.
Substituting (\ref{hor-mass-rad1a33}) and (\ref{T22g}) into (\ref{heat-capacity11}), we obtain the heat capacity as follows:
\begin{align}
\label{heat-cap1a33}
H_+ = \frac { \left( 3 c_4{r_+}^5 + {r_+}^3 + 2 c_4 \right) \pi {r_+}^2}{c_4{r_+}^5 - 2M\left(r_+\right) {r_+}^2 - 6c_4}
= - \frac { \left( 3 c_4{r_+}^5 + {r_+}^3 + 2 c_4 \right) \pi \,{r_+}^{2}}{ {r_+}^3 + 5 c_4} \,.
\end{align}
Equation~(\ref{heat-cap1a33}) shows that $H_+$ does not locally diverge and that the BH exhibits a phase transition.
The heat capacity is depicted in Figure~\ref{Fig:4}~\subref{fig:4c},
The heat capacity vanishes at the extremal limit $r_+\sim 60$
and negative when $r_+$ is smaller than the critical point $r_+\sim 60$
as in the standard BH.
In this sense, the black hole solution is not stable under the Hawking radiation.

\section{Stability of BHs (\ref{elm3}) }\label{S6}

The geodesic equations are given by
\begin{equation}
\label{ge3}
\frac{d^2 x^\alpha}{d\lambda^2}
+ \left\{ \begin{array}{c} \alpha \\ \beta \rho \end{array} \right\}
\frac{d x^\beta}{d\lambda} \frac{d x^\rho}{d\lambda}=0\, ,
\end{equation}
where $\lambda$ represents the affine connection parameter. The
geodesic deviation equations have the form \cite{1992ier..book.....D,Nashed:2003ee}
\begin{equation}
\label{ged333}
\frac{d^2 \epsilon^\sigma}{d\lambda^2}
+ 2\left\{ \begin{array}{c} \sigma \\ \mu \nu \end{array} \right\}
\frac{d x^\mu}{d\lambda} \frac{d \epsilon^\nu}{d\lambda}
+ \left\{ \begin{array}{c} \sigma \\ \mu \nu \end{array} \right\}_{,\, \rho}
\frac{d x^\mu}{d\lambda} \frac{d x^\nu}{d\lambda}\epsilon^\rho=0\, ,
\end{equation}
where $\epsilon^\rho$ is the four-vector deviation.
Substituting (\ref{ge3}) and (\ref{ged333})
into (\ref{met}), we obtain
\begin{equation}
\label{ges}
\frac{d^2 t}{d\lambda^2}=0\, , \quad \frac{1}{2} f'(r) \left(
\frac{d t}{d\lambda}\right)^2 - r\left( \frac{d \phi}{d\lambda}\right)^2=0\, , \quad
\frac{d^2 \theta}{d\lambda^2}=0 \, ,\quad \frac{d^2 \phi}{d\lambda^2}=0\, ,
\end{equation}
and for the geodesic deviation BH (\ref{met}) gives
\begin{align}
\label{ged11}
& \frac{d^2 \epsilon^1}{d\lambda^2} +h_1(r)h'(r) \frac{dt}{d\lambda}
\frac{d \epsilon^0}{d\lambda} -2r h_1(r) \frac{d \phi}{d\lambda}\frac{d \epsilon^3}{d\lambda}
+\left[ \frac{1}{2} \left( h'(r)h'_1(r)+h_1(r) h''(r)
\right)\left( \frac{dt}{d\lambda} \right)^2-\left(h_1(r)+rh'_1(r)
\right) \left( \frac{d\phi}{d\lambda}\right)^2 \right]\epsilon^1=0\, ,
\nonumber\\
& \frac{d^2 \epsilon^0}{d\lambda^2} + \frac{h'_1(r)}{h_1(r)} \frac{dt}{d\lambda}
\frac{d \zeta^1}{d\lambda}=0\, ,\quad \frac{d^2 \epsilon^2}{d\lambda^2}
+\left( \frac{d\phi}{d\lambda}\right)^2 \epsilon^2=0\, , \quad
\frac{d^2 \epsilon^3}{d\lambda^2} + \frac{2}{r} \frac{d\phi}{d\lambda}
\frac{d \epsilon^1}{d\tau}=0\, ,
\end{align}
where $h(r)$ and $h_1(r)$ are defined from Eq.~(\ref{met}) and
$'$ is derivative w.r.t. radial coordinate $r$.
The use of a circular orbit gives
\begin{equation}
\label{so}
\theta= \frac{\pi}{2}\, , \quad
\frac{d\theta}{d\lambda}=0\, , \qquad \frac{d r}{d\lambda}=0\, .
\end{equation}
Using Eq.~(\ref{so}) in Eq.~(\ref{ges}) yields
\begin{equation}
\left( \frac{d\phi}{d\lambda}\right)^2= \frac{f'(r)}{r[2h(r)-rh'(r)]}, \quad \left( \frac{dt}{d\lambda}\right)^2= \frac{2}{2h(r)-rh'(r)}\, .
\end{equation}

\begin{figure}[ht]
\centering
\subfigure[~Stability of the BH (\ref{sol3}) for $M=0.1$ and $c_5=1$]{\label{fig:11b}
\includegraphics[scale=0.25]{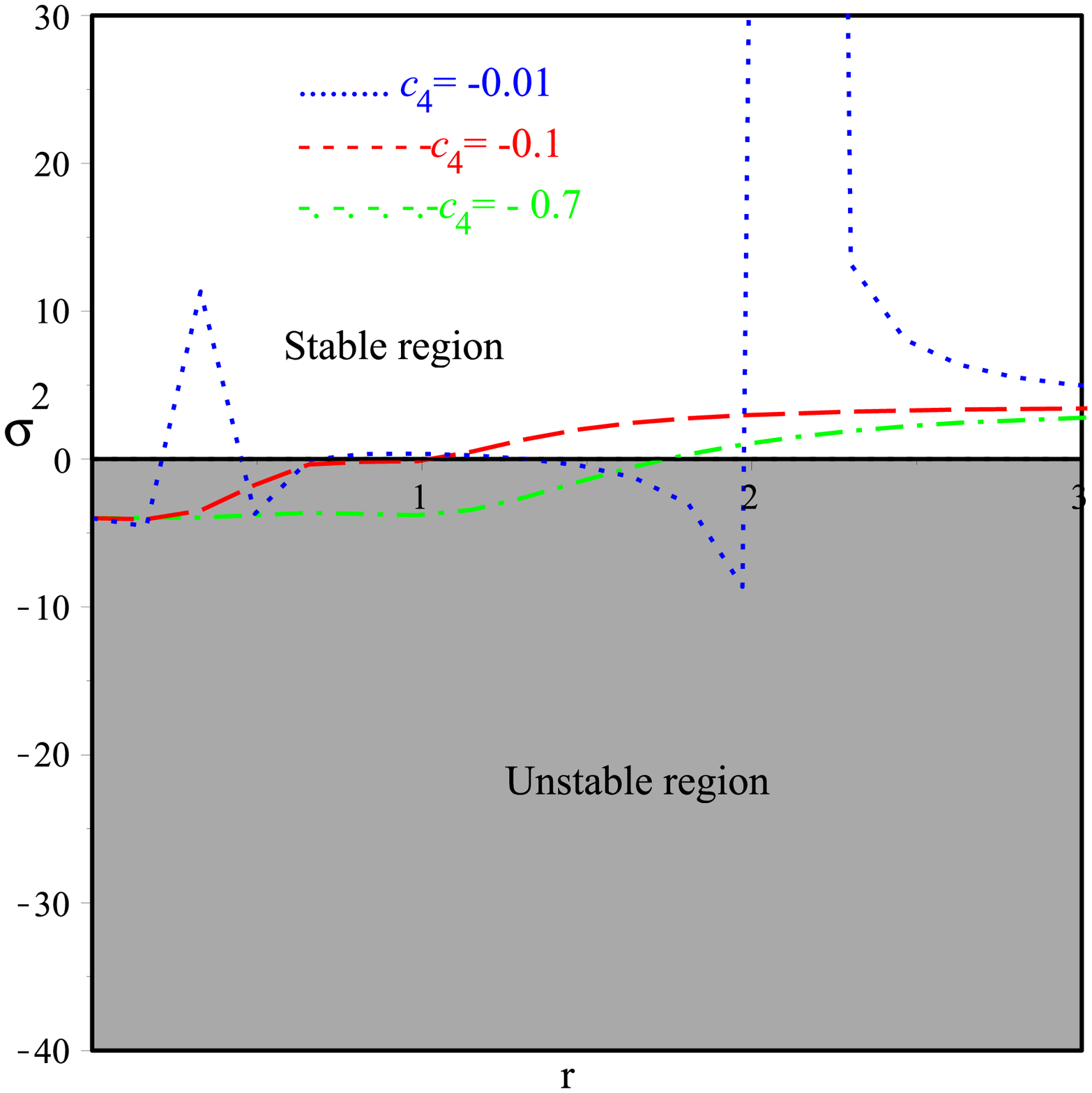}}
\subfigure[~Stability of the BH (\ref{sol3}) for $M=1$ and $c_5=1$]{\label{fig:11c}
\includegraphics[scale=0.25]{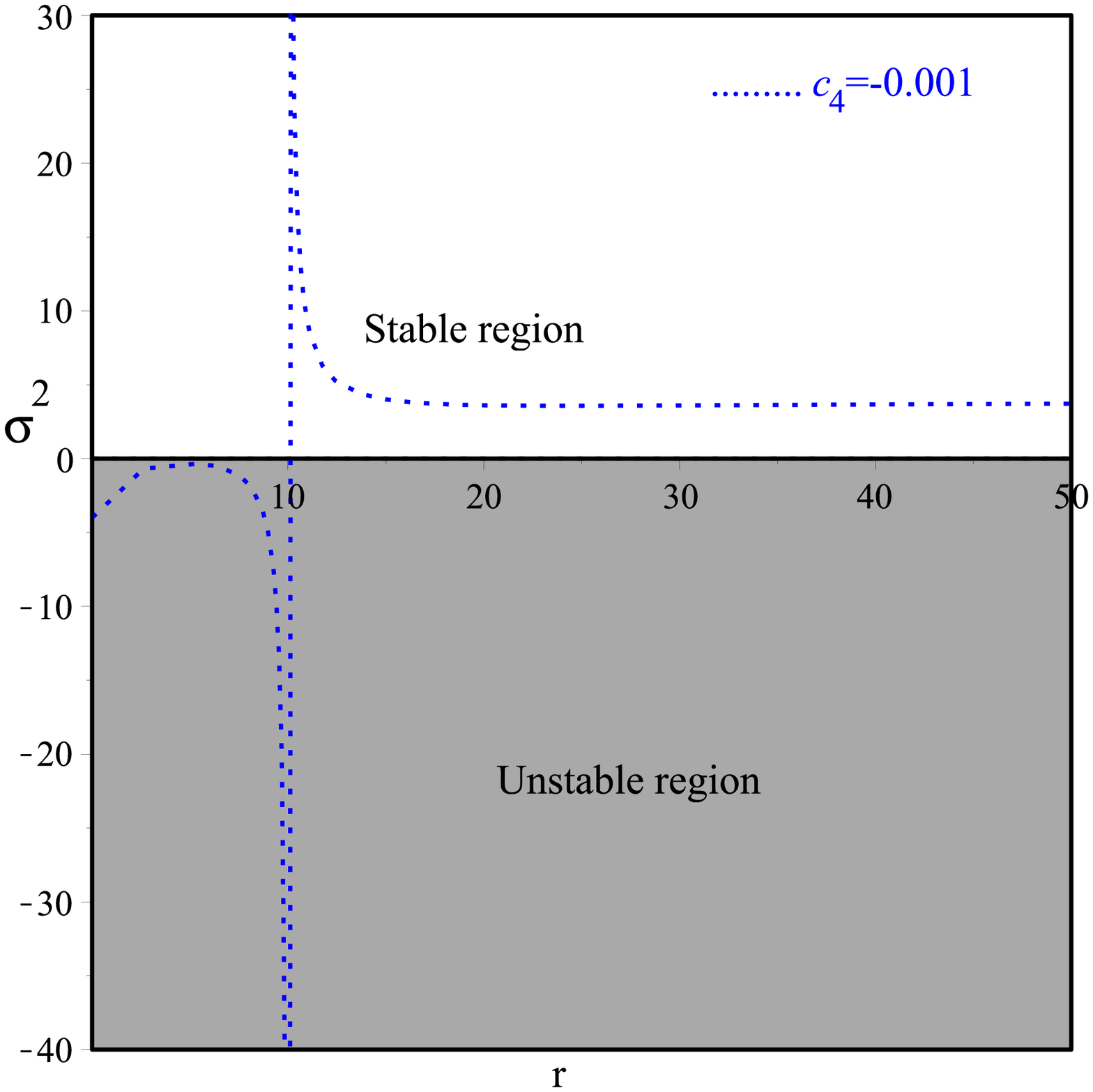}}
\caption{ {Plot of Eq.~(\ref{stab1}) against coordinate $r$ for (\ref{sol3}).}}
\label{Fig:11}
\end{figure}

The equations in (\ref{ged11}) can have the following form
\begin{align}
\label{ged22}
& \frac{d^2 \epsilon^1}{d\phi^2} +h(r)h'(r) \frac{dt}{d\phi} \frac{d \epsilon^0}{d\phi}
 -2r h_1(r) \frac{d \epsilon^3}{d\phi} +\left[ \frac{1}{2} \left[h'^2(r)+h(r) h''(r)
\right]\left( \frac{dt}{d\phi}\right)^2-\left[h(r)+rh'(r)
\right] \right]\zeta^1=0\, , \nonumber\\
& \frac{d^2 \epsilon^2}{d\phi^2}+\epsilon^2=0\, , \quad
\frac{d^2 \epsilon^0}{d\phi^2} + \frac{h'(r)}{h(r)}
\frac{dt}{d\phi} \frac{d \epsilon^1}{d\phi}=0\, , \qquad \frac{d^2 \epsilon^3}{d\phi^2}
+ \frac{2}{r} \frac{d \epsilon^1}{d\phi}=0\, .
\end{align}
 From the second equation of (\ref{ged22}) we can show that we have a simple harmonic
motion, i.e., the stability condition of plane $\theta=\pi/2$ providing that the rest of the equations in (\ref{ged22}) have solutions as follows
\begin{equation}
\label{ged33}
\epsilon^0 = \zeta_1 \e^{i \sigma \phi}\, , \quad
\epsilon^1= \zeta_2\e^{i \sigma \phi}\, , \quad \mbox{and} \quad
\epsilon^3 = \zeta_3 \e^{i \sigma \phi}\, ,
\end{equation}
where $\zeta_1$, $\zeta_2$ and $\zeta_3$ are constants.
Using Eq.~(\ref{ged33}) in (\ref{ged22}), the stability condition for spacetime (\ref{met}) is:
\begin{equation}
\label{con111}
\frac{3hh_1 h'-\sigma^2h h'-2rh_1h'^{2}+rh_1 h h'' }{h h_1'}>0\, .
\end{equation}
Equation~(\ref{con111}) has the following solution
\begin{equation}
\label{stab1}
\sigma^2= \frac{3h h_1 h''-2rh_1h'^{2}+rh h_1 h'' }{h^2 h'_1{}^2}>0\, .
\end{equation}
Using the values of $h(r)$ and $h_1(r)$ given  (\ref{sol3}),
we calculate the stability condition of the two BH solutions
and depict them in Figure~\ref{Fig:11} for the three cases of metric (\ref{met}) given by
(\ref{elm3}) using particular values of the model.

\section{Discussion}\label{S66}
Recently, most researchers trust the presence of dark matter on the account of the data coming from observations of astrophysics and cosmology.
To investigate this issue, researchers have assumed two trends to explain the dark matter:
The first trend is to modify the equation of motions of Einstein's
GR and the second one is to modify the standard model by constructing new particle species.
Many scientific papers have investigated that these two trends are the same \cite{Sebastiani:2016ras}.
Moreover, it is well--known that every modified gravitational theory has new degrees of freedom plus the usual massless graviton of Einstein's GR.
As we investigated in Section~\ref{S1} that the mimetic gravity aims to imitate DM, and it is a good candidate to investigate the presence of cold dark matter.
Thus, it is serious to check the mimetic gravitational theory in the astrophysics domain, by investigating
possible new BH solutions considering the Lagrange multiplier and mimetic potential terms.

To accomplish this study, we gave the field equations of mimetic gravitational theory coupled with the mimetic potential,
$V(r)$ and Lagrange multiplier, $\lambda(r)$, and applied them to a four-dimensional spherically symmetric
space-time having two unknown functions, $h(r)$ and $h_1(r)$, of the radial coordinate $r$.
We divided the study of the resulting non-linear differential equation into three cases:
Case I: $V(r)=\lambda(r)=0$, Case II: $V(r)=\mathrm{constant}$ and $\lambda(r) \neq 0$, and Case III: $V(r)$ and $\lambda(r)$ are not vanishing.
We show that the first case is not different from the Einstein GR which is consistent with the studies presented in the literature \cite{Nashed:2018qag}.
In the second case, the solution does not describe BH because there is no horizon. Interesting, however, the spacetime in the second case
includes the region of the Euclidian signature or the region with two times although the solution could not be physical.
In the third case, we derived non-trivial forms of the mimetic potential, Lagrange multiplier, and the metric potentials.
The BH of this case is different from those of GR and the main source of this difference comes from the non-trivial forms of the mimetic potential, and the Lagrange multiplier.
The asymptote form of this BH is AdS/dS and the source of this AdS/dS comes from the mimetic potential whose asymptote behaves like a cosmological constant.
In the third case, the spacetime has three horizons in general.

We calculated the invariants i.e., the Kretschmann $K=R_{\mu \nu \alpha \beta}R^{\mu \nu \alpha \beta}$,
the Ricci tensor squared $R_{\alpha \beta}R^{\alpha \beta}$,
and the Ricci scalar $R$, to investigate the possible singularities of the BH solutions derived from third case
and showed that all the BH solutions had true singularities at $r = 0$.
We explained that mimetic gravity coupled with mimetic potential and Lagrange multiplier produces softer singularities compared with those of GR BHs.
Moreover, we studied the energy conditions related to the third case and showed that the SEC of the second case is violated however,
all the energy conditions of the third case are verified.

We also studied the thermodynamics for the third case and find that the Hawking temperature and the heat capacity vanish in the extremal limit.
Finally, we also studied the stability of the BHs of the second and third cases using the geodesic deviation and derive the stability conditions analytically and presented it graphically.

\begin{acknowledgments}
This work is supported by the JSPS Grant-in-Aid for Scientific Research (C)
No. 18K03615 (S.N.).
\end{acknowledgments}

%

\end{document}